\documentclass[preprint,aps,nofootinbib]{revtex4}
\usepackage{array}
\usepackage{booktabs}
\usepackage{tabu}
\usepackage{dcolumn}
\usepackage{amsmath}
\usepackage{amsfonts}
\usepackage{amssymb}
\usepackage{graphicx}
\usepackage{subfigure}
\usepackage{graphicx}% Include figure files
\usepackage{dcolumn}% Align table columns on decimal point
\usepackage{bm}% bold math
\usepackage{xcolor}

\def\be{\begin{equation}}
  \def\ee{\end{equation}}
\def\bea{\begin{eqnarray}}
\def\eea{\end{eqnarray}}
\def\f{\frac}
\def\n{\nonumber}
\def\l{\label}
\def\p{\phi}
\def\o{\over}
\def\R{\rho}
\def\pa{\partial}
\def\om{\omega}
\def\na{\nabla}
\def\P{\Phi}
\begin{document}

\title{Brane inflation: swampland criteria, TCC, and reheating predictions}% Force line breaks with \\

\author{Abolhassan Mohammadi$^a$}
\email{a.mohammadi@uok.ac.ir;abolhassanm@gmail.com}
  \author{Tayeb Golanbari$^a$}
  \email{t.golanbari@uok.ac.ir; t.golanbari@gmail.com}
   \author{Salah Nasri$^{b, c}$}
   \email{snasri@uaeu.ac.ae}
           \author{Khaled Saaidi$^a$}
        \email{ksaaidi@uok.ac.ir}
  \affiliation{
$^a$Department of Physics, Faculty of Science, University of Kurdistan,  Sanandaj, Iran.\\
$^b$Department of physics, United Arab Emirates University, Al-Ain, UAE.\\
$^c$ The Abdus Salam International Centre for Theoretical Physics, Strada
Costiera 11, I-34014, Trieste, Italy.
}
\date{\today}% It is always \today, today,

\def\be{\begin{equation}}
  \def\ee{\end{equation}}
\def\bea{\begin{eqnarray}}
\def\eea{\end{eqnarray}}
\def\f{\frac}
\def\n{\nonumber}
\def\l{\label}
\def\p{\phi}
\def\o{\over}
\def\R{\rho}
\def\pa{\partial}
\def\om{\omega}
\def\na{\nabla}
\def\P{\Phi}
%\nofiles

%=============================================================%
%=============================================================%
%============== Abstract =======================================%
%=============================================================%
%=============================================================%
\begin{abstract}

We consider inflation in a five -dimensional space time with the inflaton field confined to live on a brane world.  In this scenario, we study different types of potentials for the inflaton, discuss their observational consequences, and compare with  data. We find that some class of potentials are in good agreement with observation and  that the  value of the inflaton field can be sub-Planckian. Moreover, we investigate the swampland criteria in this scenario and determine the consistency of the model with the conjectures. Doing so, we could determine models that simultaneously satisfy both observational data and swampland criteria. More constraints are applied by studying the reheating phase where the acceptable range for the reheating temperature imposes some bounds on the models. As the last step, the result of trans-Planckian censorship conjecture for the model is considered where it is shown the constraint of TCC will be very strong and it could be used to applied limit on the brane tension.

%The scenario of inflation is being studied in the frame of brane gravity where our universe is limited to a four dimensional spacetime (brane) which is embedded in five-dimensional %spacetime (bulk). The inflaton is assumed to live on the brane. The scenario is considered for different types of potential and by comparing the results with observational data, the free parameters of the model are determined. It is specified that for some potential we could have a good consistency with data, and for some other not. Also, it is determined that they give a good energy range for inflation and the values of the scalar field could be smaller than the Planck mass. Then, by using the result obtained in the first step, the recent proposed swampland criteria are investigated and it is found out that the model properly is in agreement with the criteria.
\end{abstract}
%\pacs{............}
%Use showkeys class option if keyword
\maketitle

%%%%%%%%%%%%%%%%%%%%%%%%%%%%%%%%%%%%%%%%%%%%%%%%%%%%%%%%%%%%%%%%%
%%%%%%%%%%%%%%%%%%%%%%%%%%%%%%%%%%%%%%%%%%%%%%%%%%%%%%%%%%%%%%%%%%
%%%%%%%%%%%%%%%%%%%%%%%%%%%%%%%%%%%%%%%%%%%%%%%%%%%%%%%%%%%%%%%%%%%%
%%%%%%%%%%%%%%%%%%%%%%%%%%%%%%%%%%%%%%%%%%%%%%%%%%%%%%%%%%%%%%%%%%%%
%============  Sec.I (Introduction)  =======================================
%%%%%%%%%%%%%%%%%%%%%%%%%%%%%%%%%%%%%%%%%%%%%%%%%%%%%%%%%%%%%%%%%%%%
%%%%%%%%%%%%%%%%%%%%%%%%%%%%%%%%%%%%%%%%%%%%%%%%%%%%%%%%%%%%%%%%%%%%%
%%%%%%%%%%%%%%%%%%%%%%%%%%%%%%%%%%%%%%%%%%%%%%%%%%%%%%%%%%%%%%%%%%%%%
%%%%%%%%%%%%%%%%%%%%%%%%%%%%%%%%%%%%%%%%%%%%%%%%%%%%%%%%%%%%%%%%%%%%
\section{Introduction}
The inflationary scenario is known as one of the best candidate for describing the very early universe which has been strongly supported by the observational data \cite{Planck:2013jfk,Ade:2015lrj,Akrami:2018odb}. Since the first proposal of the scenario \cite{starobinsky1980new,Guth:1980zm,albrecht1982cosmology,linde1982new,linde1983chaotic} many inflationary model have been introduced such as non-canonical inflation \cite{Barenboim:2007ii,Franche:2010yj,Unnikrishnan:2012zu,Gwyn:2012ey,Rezazadeh:2014fwa,Cespedes:2015jga,Stein:2016jja,Pinhero:2017lni}, tachyon inflation \cite{Fairbairn:2002yp,Mukohyama:2002cn,Feinstein:2002aj,Padmanabhan:2002cp}, DBI inflation \cite{Spalinski:2007dv,Bessada:2009pe,Weller:2011ey,Nazavari:2016yaa,Amani:2018ueu,Mohammadi:2018zkf}, G-inflation \cite{maeda2013stability,abolhasani2014primordial,alexander2015dynamics,tirandari2017anisotropic}, warm inflation \cite{berera1995warm,berera2000warm,taylor2000perturbation,hall2004scalar,BasteroGil:2004tg,Bastero-Gil:2016qru,Rosa:2018iff,Bastero-Gil:2019gao,Sayar:2017pam,Akhtari:2017mxc,Sheikhahmadi:2019gzs}, in which the most common picture is that inflation is drives by a scalar field which slowly rolls down to minimum of its potential \cite{Riotto:2002yw,Baumann:2009ds,Weinberg:2008zzc,Lyth:2009zz}. \\
After inflation the universe is cold and almost empty of particles. Then, a mechanism is required to warm up the universe and fill it with particles. The mechanism is known as (p-)reheating \cite{Abbott:1982hn,Albrecht:1982mp,Dolgov:1982th,Dolgov:1989us,Traschen:1990sw,Shtanov:1994ce,Kofman:1994rk,Kofman:1997yn,Bassett:2005xm,Allahverdi:2010xz,Amin:2014eta} describing an energy transfer from scalar field to other field leading to particle production. The produced particles interact and thermalize the universe and allow a smooth transition to radiation dominant phase. The reheating temperature should on one side be large enough to recover the successful hot big bang nucleaosynthesis ($T > 1 \; {\rm MeV}$) and also small enough to avoid the reproduction of any unwanted particle ($T < 10^{9-10} \; {\rm GeV}$) \cite{Kofman:1994rk,Kofman:1997yn,Bassett:2005xm,Allahverdi:2010xz,Amin:2014eta}. Reheating is inseparable part of (cold) inflation model, and any inflation model without reheating is incomplete.  \\
The standard model of inflation has been generalized in different ways which one of them is the inflationary scenario in modified gravity models where the brane gravity model is known as one of the interesting generalized theory of gravity. The brane theory of gravity is a higher dimensional model of gravity which has been inspired from M-theory. The first model of brane world was introduced by Randall and Sundrum (RS) in 1999 where the main motivation of the model was to find a solution for the Hierarchy problem between electroweak scale and Planck scale \cite{Randall:1999ee,Randall:1999vf}. The general picture is that all standard particles are confined to a four-dimensional space-time (brane) and only gravity could propagates in higher dimension. In other words, our universe is a three brane embedded in five-dimensional space-time which is called bulk. The model introduces an interesting and novel feature in the evolution equation. The Firedmann equation in brane world gravity includes both quadratic and linear terms of the energy density while in four-dimensional cosmology there is only linear term. The quadratic term of the energy density dominates over the linear term in the high energy regime (where energy density is larger than the brane tension, i.e. $\rho \gg \lambda$). Consequently, the Hubble parameter in this regime is proportional to the energy density, $H \propto \rho$ and it is no longer proportional to $H \propto \sqrt{\rho}$ \cite{maartens2000chaotic,golanbari2014brane,Mohammadi:2020ftb,Banerjee:2017lxi,Elizalde:2018rmz,Paul:2018jpq}.
A theoretical constraint on the inflationary models has been recently proposed which is known as the swampland criteria \cite{Obied:2018sgi,Garg:2018reu,Ooguri:2018wrx}. The origin of these criteria stands in string theory where they are realized as a measure to recognize the consistence low-energy effective field theory (EFT) from the inconsistence ones. It includes two conjectures:
%After that, the swampland criteria is considered for each case to find out whether the scenario could successfully pass them. String theory which is known as one of the candidate of quantum gravity, includes two region of the effective field theory (EFT) called landscape and swampland. The landscape consists consistence low-energy EFT that could properly formulate the quantum gravity. However, there is another larger region, swampland, that surrounds the landscape. The low-energy EFTs, which live on the swampland, are in contradiction with string theory. There have been huge effort to introduce some conditions and criteria to separate these two types of EFT which was a direct results of our natural desire for having consistence EFT. The recent studies on effective field theory (EFT) and string theory leads to two swampland criteria:
I) There is an upper bound on the field range, i.e. \cite{Obied:2018sgi,Garg:2018reu,Ooguri:2018wrx}
${\Delta\phi / M_p} < c_1$ where $c_1$ is of order of unity, which rise from this belief that the effective Lagrangian in the EFT is valid only for a finite radius;
II) putting an upper bound on the gradient of the potential of the field of any EFT, i.e. \cite{Obied:2018sgi}
$M_p \; { |V'| / V} \geq c_2$ or the refined version of this conjecture, given by \cite{Garg:2018reu,Ooguri:2018wrx}
$M_p^2 \; { V'' / V} \geq -c_3$ where $c_3$ is positive. The refined conjecture states that the potential must be sufficiently tachyonic. Also, the most recent studies determines that $c_2$ and $c_3$ could be even of order of $\mathcal{O}(0.1$) \cite{Kehagias:2018uem,Ooguri:2018wrx}.
In the first look, the second criterion is in direct tension with the slow-roll inflation where the slow-roll parameter $\epsilon_{\phi} = M_p^2 (V' / V)^2$ must be smaller than one. In general, these two criteria rule out some of the inflationary models, however, the recent studies \cite{Kehagias:2018uem,Das:2018rpg,Kinney:2018kew,Matsui:2018bsy,Lin:2018rnx,Dimopoulos:2018upl,Kinney:2018nny,Geng:2019phi,Brahma:2018hrd,Brahma:2019iyy,Wang:2019eym,Odintsov:2020zkl,Odintsov:2018zai,Rasheed:2020syk,Brahma:2020cpy} have determined that some non-standard models of inflation might still survive these two criteria, in which the brane inflation could be one of them. \\
The more recent conjecture is the trans-Planckian censorship conjecture (TCC) proposed in \cite{Bedroya:2019snp}. The conjecture states that for any consistent theory of quantum gravity it is absolutely impossible that a mode that was trans-Planckian never can cross the Hubble horizon. This situation never happens for a model like big bang cosmology where the mode never cross the horizon. However, for inflationary phase this conjecture might lead to some serious outcomes. The TCC for standard inflation results in some strong condition on the energy scale and tensor-to-scalar ratio \cite{Bedroya:2019tba}, and the condition get more stronger for brane inflation \cite{Mohammadi:2020ctd}. The investigation were performed by this assumption that the Hubble parameter is constant during inflation and the reheating phase occurs very fast.  \\
%The inflation is assumed to be happened in the energy scale below the Planck scale. Therefore it is believed that the scenario could be described by a low-energy EFT. It is our interest to build it based on a consistent EFT that lives on landscape and not swampland. Due to this, in the first sight, it is the second swampland criterion that might put the inflationary models at risk because the second criterion implies that the first slow-roll parameter, i.e. $\epsilon_{\phi} = M_p^2 (V' / V)^2,$ and the tensor-to-scalar ratio, $r = 16\epsilon_{\phi} > 8c_2^2,$  should not be small. However, according to the fundamental assumption of the inflationary scenario, the first slow-roll parameter $\epsilon$ must be smaller than one which is in direct tension with the second swampland criterion. The desire for satisfying these two criteria could roll out some of the inflationary models, however the recent studies \cite{Kehagias:2018uem,Das:2018rpg,Kinney:2018kew,Matsui:2018bsy,Lin:2018rnx,Dimopoulos:2018upl,Kinney:2018nny,Geng:2019phi} have determined that some non-standard models of inflation might still survive these two criteria, in which the brane inflation could be one of them. \\
The main reasons that motivates us to consider the inflationary scenario in the frame of RSII brane gravity model are two folds: observational and theoretical consistency. First, due to interesting feature of the Friedmann equation in the brane world model which is expected to lead to some novel conclusions. The scenario is studied for different well-known potentials, and the free parameters of the model are determined by comparing the results with observational data. In this regard, our method is different from the previous studies where instead of testing the results of the model for two or three sets of the constant parameters of the model, we find a parameter space in which every point is consistent with the data. The observational data, during the past years, is getting better and there are chances that some of the potentials be throwing out due to their inconsistency with data. Considering the consistency with the swampland criteria is another motivation for the present work. There is a growing interest to find the inflationary models which simultaneously agree with observational data and swampland criteria. Then, the model prediction about the reheating temperature is investigated which leads to more constraint on the parameters. Consistency with the TCC, as the most recent conjecture, will be another topic that would be interesting to be studied.   \\

%On the other hand, during past years, the observational data is getting better and there are chance that some of the potential be throwing out due to their inconsistency with data. In this case, the chance of the success in the frame of brane gravity is higher. The second reason goes to the swampland criteria, in which it sounds that the inflation in the frame of brane world have a better chance to also satisfy the two swampland criteria. \\

The paper is organized as follows: After the introduction, the main dynamical equations of the model are presented in Sec.II. In Sec.III, the slow-roll parameters are introduced for a general form of the potential, and the perturbations parameters are described in terms of the potential. Next, in Sec.IV we are going to consider the consistency of the model with data for different well-known types of the potential, then try to find out the consistency of the result with the swampland criteria. The model prediction about the reheating temperature is considered in Sec.V, where it is realized that to have an acceptable value for the temperature, the parameters are required to be restricted more. In Sec. VI, some note about the TCC are presented and the applied condition on the model is discussed. The results will be summarize and discussed in Sec.VII.

%=============================================================%
%=============================================================%
%=============================================================%
%=============================================================%
%============== Section 2 =======================================%
%=============================================================%
%=============================================================%
%=============================================================%
%=============================================================%
\section{The Model}
Our study will be  limited to Randall-Sundrum II brane gravity model, with the following action
\begin{equation}\label{action}
S_5 = \int d^5x \sqrt{-g} \Big( {M_5^3 \over 2}\mathcal{R} - \Lambda_5 \Big) + \int d^4x \sqrt{-q} \Big( L_{\rm b} - \lambda \Big),
\end{equation}
where  $\mathcal{R}$ is the Ricci scalar,  $\Lambda_5$  the five-dimensional cosmological constant
%which fills the bulk and
,  $M_5$ stands for five-dimensional Planck mass,  and $q_{\mu\nu}$  the induced metric on the brane which is related to the  five-dimensional metric $g_{AB}$ by the relation $g_{AB}=q_{AB} + n_A n_B$, where $n^A$ is a unit normal vector. $L_{\rm b}$ indicates the Lagrangian of matter that has confined on the brane and $\lambda$ is the brane tension. By taking variation of the above action with respect to the metric %and applying the $Z_2$-symmetry,
we obtain the  the field equation of motion % the model is read as
\begin{equation}\label{fieldEq}
G_{\mu\nu} = -\Lambda_4 g_{\mu\nu} + \left({8\pi\over M_{4}^2}\right) \tau_{\mu\nu}  + \left({8\pi\over
M_5^3}\right)^2 \Pi_{\mu\nu} - E_{\mu\nu}\ ,
\end{equation}
with
\begin{eqnarray}
\Lambda_4 & = & {4\pi \over M_5^3} \left( \Lambda_5 + {4\pi\over3M_5^3} \lambda^2 \right), \label{4CC} \nonumber \\
M_4^2 &=& \frac{3}{ 4\pi} \frac{M_5^6}{\lambda}\nonumber \\
E_{\mu\nu} & = & C_{MRNS}~n^M n^N q_{~\mu}^{R}~ q_{~\nu}^{S} , \nonumber \\
\tau_{\mu\nu} & = & -2 {\delta L_{\rm b} \over \delta g^{\mu\nu}} + g_{\mu\nu}L_{\rm b} , \nonumber \\
\Pi_{\mu\nu} & = & -\frac{1}{4} \tau_{\mu\alpha}\tau_\nu^{~\alpha} +\frac{1}{12}\tau\tau_{\mu\nu} +\frac{q_{\mu\nu}}{8}\tau_{\alpha\beta}\tau^{\alpha\beta} - \frac{q_{\mu\nu}}{24} \tau^2  . \nonumber
\end{eqnarray}
Here $M_4$ is   the effective four-dimensional Planck mass,  $\Lambda_4$  the  cosmological constant on the brane is defined by $\Lambda_4$ which is a combination of the five-dimensional cosmological constant and the tension of the  brane,  $E_{\mu\nu}$ is the projection of the five-dimensional  Weyl tensor $C_{MRNS}$ on the brane, and $\tau_{\mu\nu}$ is the brane energy momentum tensor.   Note that  both the linear and quadratic terms contribute to the    effective four-dimensional  energy-momentum tensor. \\

Assuming  the homogeneity and isotropy of  the universe  and  a  spatially flat five-dimensional Friedmann–Lemaitre–Robertson–Walker (FLRW) metric, defined as
\begin{eqnarray} \label{FLRW}
ds^2_5 = -dt^2 + a^2 \delta_{ij} dx^i dx^j + dy^2  ,
\end{eqnarray}
where $\delta_{ij}$ is a maximally symmetric three-dimensional metric and $y$ denotes  the fifth coordinate, the corresponding  Friedmann equation reads
\begin{equation}
H^2={\Lambda_4 \over 3} + \left( {8\pi \over 3M_4^2} \right) \; \rho + \left({4\pi\over 3M_5^3}\right)^2 \; \rho^2  +{\mathcal{C} \over a^4} . \nonumber
\end{equation}
The last term on the right hand side of the above equation arises from the term $E_{\mu\nu}$,  which  descibes the influence of the bulk graviton on the brane evolution, and is known as the dark radiation. Because  it scales as $a^{-4}$, the dark radiation  gets rapidly  diluted  during inflationary phase, and hence  can be neglected. Also,  we will set $\Lambda_4 = 0$ as  in the original  RS model. Therefore, the Friedmann equation is rewritten as
\begin{equation}\label{Friedmann}
H^2 = {8\pi \over 3M_4^2} \rho \left( 1+ {\rho \over 2\lambda} \right) ,
\end{equation}
%The four-dimensional Planck mass $M_4$ is related to the five-dimensional one through $M_4^2 = {3 \over 4\pi} {M_5^6 \over \lambda}$.
In  the high energy region, there the contribution  of  the term  quadratic in the energy density is  dominant in the expression of the Hubble parameter, where as  in the regime  where  $\rho \ll \lambda$, the Friedmann equation  the usual form of standard cosmology. Since  standard  cosmology  is very successful in describing the evolution of  the universe from the time of  nucleosynthesis,  it requires  the brane tension as $\lambda \geq 1 {\rm MeV^4}$,  leading to the five-dimensional Planck mass $M_5 \geq 10 {\rm TeV}$ \cite{Cline:1999ts,Germani:2001du}. Moreover, the Newtonian law of gravity receives a correction of order $M_5^6/\lambda^2 r^2$, which should be small on scales larger than $r \geq 1 {\rm mm}$, and consequently yields to  the stronger  constraint  $M_5 \geq 10^5 {\rm TeV}$ \cite{Germani:2001du}. There are  also various  astrophysical implications which  set  strong limit on the brane tension  $\lambda \geq 5 \times 10^8 {\rm MeV^4}$ (see  \cite{Germani:2001du}).   \\

The matter confined to  the brane satisfy the same  energy conservation equation as  in standard cosmology, i.e
\begin{equation}\label{conservation}
  \dot{\rho} + 3 H (\rho+p)=0.
\end{equation}
Using this equation and  taking the time derivative of Eq.\eqref{Friedmann}, we obtain the second  Friedmann equation
\begin{equation}\label{SFriedmann}
  \dot{H} = {-4 \pi \over M_4^2} \; \left( 1 + {\rho \over \lambda} \right) \; (\rho + p).
\end{equation}

%=============================================================%
%=============================================================%
%=============================================================%
%=============================================================%
%============== Section 3 ====================================%
%=============================================================%
%=============================================================%
%=============================================================%
%=============================================================%
\section{Brane inflation}
%Inflation is known as the phase of an extreme expansion that the universe undergoes in very short time.
We assume  the inflaton is scalar field  living on the brane  and has  the  energy density and pressure $\rho = {\dot{\phi}^2 \over 2 } + V(\phi)$ and $p= {\dot{\phi}^2 \over 2 } - V(\phi)$, respectively, which  % , which drive inflation,  a scalar field, called inflaton, which dominates the %universe in the time.
%In the present work, the scalar field lives on the brane with the following energy density and pressure
%\begin{equation}\label{energypressure}
%\rho = {\dot{\phi}^2 \over 2 } + V(\phi), \qquad p= {\dot{\phi}^2 \over 2 } - V(\phi)
%\end{equation}
 is governed by  the equation of motion
\begin{equation}\label{EoM}
\ddot{\phi} + 3H \dot{\phi} + V'(\phi) = 0.
\end{equation}
where $V(\phi)$ is the potential of the inflaton. The common picture for the universe is that the scalar field slowly rolls down toward the minimum of its potential. During  this slow-rolling phase, the scalar field
yields very small kinetic energy which can be  neglected  compared to its  potential energy. Also,  it is assumed that the  term $\ddot{\phi}$ is much smaller than the friction term $H \dot{\phi}$ and the slope of the potential $V'$. These assumptions  are known as the slow-roll conditions and are described by  the smallness of the slow-roll parameters:% in which the two important of these parameters are
\begin{equation}\label{SRPH}
\epsilon = {-\dot{H} \over H^2}, \qquad \eta = {-\ddot{\phi} \over H \dot{\phi}}
\end{equation}
With these parameters, the dynamical equations of the model could be rewritten as
\begin{eqnarray}
  H^2 & = & {8\pi \over 3M_4^2} \; V(\phi) \; \left( 1+ {V(\phi) \over 2\lambda} \right), \label{FriedmannV} \\
  \dot{H} & = & {-4 \pi \over M_4^2} \; \left( 1 + {V(\phi) \over \lambda} \right) \; \dot{\phi}^2, \\
  3 H \dot{\phi} & = & -V'(\phi). \label{phidot}
\end{eqnarray}
Using these equations,  we can express the slow-roll parameters  in terms of the potential  and its derivatives as
\begin{eqnarray}
  \epsilon &=& {M_4^2 \over 16\pi} \; \left( {V'(\phi) \over V(\phi)} \right)^2
  {4\lambda \big(\lambda + V(\phi) \big) \over \big( 2\lambda + V(\phi) \big)^2}, \\
  \eta &=& {M_4^2 \over 8\pi} \; {V''(\phi) \over V(\phi)} \; {2\lambda \over 2\lambda + V(\phi)}.
\end{eqnarray}

Compared to the standard cosmology, here we have a generalized Friedmann equation with some modified terms. These It is important to note that  in the high energy limit, i.e. $\rho \gg \lambda$, The quadratic term of the energy density dominates over the linear term and the Hubble parameter is proportional to the potential, in contrast to the standard cosmology where $H \propto V^{1/2}(\phi)$. For the rest of the work, we will assume  that  inflation occurs in the high energy limit, in which case  the slow-roll parameters get  the  simpler form
\begin{equation}\label{SRPV}
\epsilon = {1 \over 3} \; \left( {3 M_5^3 \over 4\pi} \right)^2 {V'^2(\phi) \over V^3(\phi)} , \qquad
\eta = {1 \over 3} \; \left( {3 M_5^3 \over 4\pi} \right)^2 {V''(\phi) \over V^2(\phi)}
\end{equation}
%To solve the problem of big-bang theory based on the scenario of inflation, the universe should inflates enough.

The   expansion of the universe during inflation  is quantified by the  number of e-fold  which describes how  long  this  exponential phase should last, and is  defined as
\begin{equation}\label{efold}
N = \int_{t_i}^{t_e} H \; dt =
-3 \left( {4\pi \over 3 M_5^3} \right)^2 \int_{\phi_i}^{\phi_e} \; {V^2(\phi) \over V'(\phi)} \; d\phi
\end{equation}
where in obtaining the second equality  Eqs.\eqref{FriedmannV} and \eqref{phidot}  have been used.%have been used.

%=============================================================%
%=============================================================%
%=============================================================%
%=============================================================%
%=============================================================%
\subsection{Cosmological perturbations}
The quantum perturbations in the inflationary scenario are divided into three types: scalar, vector, and tensor, in which the scalar perturbations are the seeds for large scale structure of the universe and tensor perturbations are known as the primordial gravitational waves. The vector perturbations are less important since they behaves as the inverse of the scale factor and rapidly diluted during inflation.  \\

Let us consider an arbitrary scalar perturbation to the background FLRW metric
\begin{eqnarray}\label{perturbedmetric}
ds^2 & = & -(1+2A)dt^2 - 2a^2(t)\nabla_i B dx^i dt \nonumber \\
 & & + a^2(t)\Big[(1-2\psi)\delta_{ij} + 2 \nabla_i\nabla_j E  \Big]dx^i dx^j .
\end{eqnarray}
where $\delta_{ij}$ is the spatial metric of the background and $\nabla_i$ stands for covariant derivative with respect to the metric. The quantity $\psi$ is called the curvature perturbations due to the fact the intrinsic curvature of the spatial hypersurface is directly related to the this parameter as $^{3}\mathcal{R} = 4 \nabla^2 \psi / a^2$. The curvature perturbation is  gauge dependent and changes under arbitrary coordinate transformation. However, the curvature perturbations in the uniform density hypersurface, given by $\zeta = \psi + {H \delta \rho \over \dot{\rho}} $ is a gauge invariant perturbation parameter. For the single scalar field inflationary models , where the perturbations can be assumed to be adiabatic, the curvature perturbation $\zeta$ is conserved and remains almost constant at  the superhorizon scale \cite{maartens2000chaotic,Wands:2000dp}. This is the most important feature of the parameter. On the spatially flat hypersurface, $\psi = 0$,  using the scalar field energy density, the gauge-invariant curvature perturbation $\zeta$ is obtained as
\begin{equation}\label{zeta}
  \zeta = {H \over \dot{\phi}} \; \delta\phi,
\end{equation}
where $\delta \phi = H / 2\pi$. Following \cite{maartens2000chaotic,Wands:2000dp}, the amplitude of the scalar perturbation is defined as $\mathcal{P}_s = 4 \langle \zeta^2 \rangle / 25$, and making use of  the slow-roll approximations  we have %it is given as [arXiv:hep-ph/9912464]
\begin{equation}\label{ps}
\mathcal{P}_s = {9 \over 25\pi^2} \; \left( {4\pi \over 3 M_5^3} \right)^6 \; {V^6(\phi) \over V'^2(\phi)}
\end{equation}
Using above relation, we obtain the scalar spectral index
\begin{equation}\label{ns}
n_s - 1 = {d \ln(\mathcal{P}_s) \over d\ln(k)} = -6 \epsilon + 2 \eta.
\end{equation}
The  derivation of the amplitude of the  tensor perturbations for this  model is a little trickier than in the standard four-dimensional cosmology since here the graviton can propagate along the fifth dimension as well. It is given by \cite{Brax:2004xh,Langlois:2000ns,Huey:2001ae}
 \begin{equation}
\mathcal{P}_g = {16\pi \over 25 \pi \; M_p^2} \left( {H \over 2\pi} \right)^2 F^2(x)
\end{equation}
where
\begin{equation}
F^2(x) = \left[ \sqrt{1+x^2} \; - \; x^2 \; \sinh^{-1}\left({1 \over x} \right) \right]^{-1/2}, \qquad
x \equiv \sqrt{3 \over 4\pi \lambda} \; M_p \; H
\end{equation}
Using  Eq.\eqref{FriedmannV} and considering   the  high energy limit,  the perturbations reads% $F^2(x) \rightarrow  3x/2$, and hence}% then $\mathcal{P}_g$ is obtained as
\begin{equation}
\mathcal{P}_g = {9 \over 50 \pi^2} \left( { 4\pi \over 3 M_5^3 } \right)^4 \; {V^3(\phi)}
\end{equation}
%{\textcolor{blue}{\bf which we made use  of Eq.\eqref{FriedmannV}  to obtain the above relation}}.
The tensor perturbation is measured indirectly through  the  tensor-to-scalar ratio
%$r = {3 \; \epsilon \over 2}$, which constrained by the Planck data  % given by
\begin{equation}\label{r}
  r = {3 \; \epsilon \over 2}.
\end{equation}
Thus far there is no evidence for such contribution  which yields   an upper limit $r < 0.064$ from the Planck data combined  with the BICEP2/Keck Array BK14 data \cite{Akrami:2018odb}. \\

Note that the general shape of the main perturbation parameters $\mathcal{P}$, $n_s$, and $r$ might seem more and less similar to the standard onse, however they are different because there is a different Hubble parameter. The main difference that is our concern is that The Hubble parameter in brane gravity depends on $\rho$, and not on $\sqrt{\rho}$. As a result, the evolution and behavior of the fluctuations, the slow-roll parameters, and in tern the behavior of the perturbation parameters are different than the onse in standard 4D gravity. Therefore, the results that we are comparing with observational data are not the onse we have in standard general relativity.

% Y.~Akrami {\it et al.} [Planck Collaboration],
  %``Planck 2018 results. X. Constraints on inflation,''
 % arXiv:1807.06211 [astro-ph.CO].

%=============================================================%
%=============================================================%
%=============================================================%
%=============================================================%
%============== Section 4 =======================================%
%=============================================================%
%=============================================================%
%=============================================================%
%=============================================================%
\section{Consistency with observation and swampland criteria}\label{ObservationSwampland}
%To consider the validity of the model, its consistency with observational data should be investigated.
In this section,  we we consider  in details  different types of inflaton potentials and for each we  determine the model parameter space that is consistent with the latest observational data \cite{Akrami:2018odb}.

%=============================================================%
%=============================================================%
%=============================================================%
%=============================================================%
%=============================================================%
\subsection{Power-law Potential}
As the first case, the power-law potential is picked out. Although the power-law potential in the standard inflation model could not be a good choice for describing the inflation, it could have a proper consistency with data in the modified theories of gravity, e.g. scalar-tensor theory of gravity. One of the main features of the potential which puts it in the center of our attention is its simplicity. Due to this fact, the potential is the first choice for considering any inflationary model. It is always desirable to have a simple model for describing a phenomenon, like inflation. The power-law potential is given by
\begin{equation}\label{Vpowerlaw}
V(\phi) = V_0 \; \phi^n,
\end{equation}
where $V_0$ and $n$ are constant. Substituting this potential into Eq.\eqref{SRPV} yields  the slow-roll parameters% are acquired as
\begin{equation}\label{SRP-Vpower}
  \epsilon = \left( {3 M_5^3 \over 4\pi } \right)^2 \; {n^2 \over 3V_0} \; {1 \over \phi^{n+2}}, \qquad
  \eta =  \left( {3 M_5^3 \over 4\pi } \right)^2 \; {n(n-1) \over 3V_0} \; {1 \over \phi^{n+2}}
\end{equation}
By setting $\epsilon = 1$, we can infer the value of  the scalar field at the end of inflation as
%To determine the value of  the scalar field at the end of inflation  we set  $\epsilon = 1$, solve for ...m and apply  it  Eq.\eqref{efold} to get
%The scalar field at the end of inflation is obtained by solving the relation $\epsilon = 1$,  and we get%which implies the end of accelerated expansion phase of the Universe. Doing so, one arrives at
\begin{equation}\label{phiend-Vpower}
 \phi_e^{n+2} = \left( {3 M_5^3 \over 4\pi } \right)^2 \; {n^2 \over 3V_0}.
\end{equation}
Applying this result to Eq.\eqref{efold} yields  the scalar field at the horizon crossing\footnote{Note  that different values of $N$ correspond  to  different horizon crossing times. The point is that fluctuations are produced constantly during inflation. Some exit the horizon at earlier times, which is indicated by larger value of $N$. Some exit the horizon at later times which is expressed by smaller values of $N$.\\
In inflationary scenario, it is assumed that the fluctuations (that we are interested in) cross the horizon and then there is about $N=55-65$ e-fold of expansion. A common way to count the expansion is that we asign $N=0$ to the end of inflation, and $N=55-65$ is associated with the time of horizon crossing (i.e. we only reverse the e-fold counting and the result will be the same). We sometimes call the latter time as the beginning of inflation. Then, $N=0$ means the end of inflation and $N=60$ (for instance) stands for the beginning of inflation. This point is clear in Eq.\eqref{phis-Vpower}, in which if one put $N=0$, the result is exactly equal to $\phi_e$, Eq.\eqref{phiend-Vpower}. The scalar field at the obeginning of inflation is obtained by puting $N=60$.}
\begin{equation}\label{phis-Vpower}
  \phi_\star^{n+2} = \left( {3 M_5^3 \over 4\pi } \right)^2 \; {n^2 \over 3V_0} \; \left[ 1 + {(n+2) \; N \over n} \right].
\end{equation}
To obtain the energy scale of inflation, one only needs to substitute the scalar field $\phi\star$ in the potential Eq.\eqref{Vpowerlaw}. Doing so, we have
\begin{equation}
V_{es} = V_0 \; \left( {3 M_5^3 \over 4\pi } \right)^{2n \over n+2} \; \left( {n^2 \over 3V_0} \; \left[ 1 + {(n+2) \; N \over n} \right] \right)^{n \over n+2}
\end{equation} 
which presents a relationship between the energy scale of inflation and number of e-fold. By the energy scale we means the value of the potential at begining of inflation, namely $N=55-65$. The energy scale of inflation is indicated by $V^\star$ which is equal to $V_{es} = V|_{N=55-65}$. \\

\begin{figure}[h]
  \centering
  \includegraphics[width=6cm]{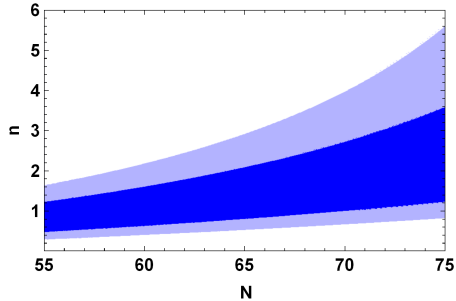}
  \caption{The  parameter space  $(n, N)$ for the power law potential   that  yield  values $(r, n_s)$ allowed by  the Planck data at the $95\%$ CL  (light blue)and the $68\%$ CL (dark blue).}\label{nN}
\end{figure}

Substituting Eq.\eqref{phis-Vpower} into  Eqs.\eqref{SRP-Vpower} , \eqref{ns}, and \eqref{r}, we get
\begin{equation}\label{SRP-VpoweN}
  \epsilon(n,N) = \left( 1 + {(n+2) \; N \over n} \right)^{-1}, \qquad
  \eta(n,N) = {(n-1) \over n} \; \left( 1 + {(n+2) \; N \over n} \right)^{-1}
\end{equation}
which is a  function of only  the power of the potential and the  number of e-folds. Consequently, %.  Hence, Then, from Eqs.\eqref{ns} and \eqref{r}, it is realized that
the scalar spectral index and the tensor-to-scalar depend on just on the  two parameters $n$ and $N$. Using the Planck $r-n_s$ diagram,  we show in  Fig.\ref{nN} the parameter space for $(n,N)$ for  which the model predictions are in agreement  withe Planck data.

%%%%%%%%%%%%%%%%%%%%%%%%%%%%%%%%%
%%%%%%%%%%%%%%%%%%%%%%%%%%%%%%%%%

Also, from the expression  of the amplitude of the scalar perturbations, Eq.\eqref{ps}, the constant $V_0$ is determined as
\begin{equation}\label{V0-power}
  V_0^{6 \over n+2} = {25 \pi^2 n^2 \mathcal{P}_s \over 9} \; \left( {3 M_5^3 \over 4\pi} \right)^6   \; \left[ {3n^2 \over 16 \pi^2} \; M_5^6 \; \left( 1 + {(n+2) N \over n} \right) \right]^{-2(2n+1) \over n+2}
\end{equation}
and depends on the values of $n$, $N$, and  the five-dimensional Planck mass. Setting $M_5 = 2 \times 10^{14} \; {\rm GeV}$, in Table.\ref{Tpower} we present the values of  $V_0$ and the energy scale $V^\star$  for different values of $n$ and $N$ taken from Fig.\ref{nN}.
%The constant $V_0$ and the energy scale $V^\star$ has been determined and presented in Table.\ref{Tpower} for different values of $n$ and $N$, taken from Fig.\ref{nN}.
Considering the constraint on the brane tension $\lambda$, stated in Sec.II, and for the same chosen value of $M_5$,  we  find that  $\rho/\lambda \sim \mathcal{O}(10^{11}-10^{18})$, which is  much larger  than unity,  and hence  in consistent  with our  assumption that  inflation occured  in the high energy regime.  \\
%%%%%%%%%%%%%%%%%%%%%%%
\begin{table}
  \centering
  \begin{tabular}{p{1cm}p{1cm}p{3.5cm}p{3cm}}
    \hline
     $n$ &  $N$ & \qquad\quad $V_0$ & \qquad \quad $V^\star$ \\
    \hline
    $1$   & $60$ & $2.08 \times 10^{55} \; {\rm GeV^3}$     & $4.57 \times 10^{65} \; {\rm GeV^4}$  \\
    $1.5$ & $65$ & $5.07 \times 10^{42} \; {\rm GeV^{5/2}}$ & $3.35 \times 10^{61} \; {\rm GeV^4}$ \\
    $2$   & $70$ & $9.03 \times 10^{29} \; {\rm GeV^{2}}$   & $2.49 \times 10^{58} \; {\rm GeV^4}$\\
    \hline
  \end{tabular}
  \caption{The constant $V_0$ and the energy scale of the inflation for different values of $(n,N)$ taken from Fig.\ref{nN} and $M_5=2 \times 10^{14} \; {\rm GeV}$}\label{Tpower}
\end{table}
%%%%%%%%%%%%%%%%%%%%%%%

%So far we  determined the model parameters  the observational data.

\begin{figure}[h]
  \centering
  \subfigure[\label{SC1A}]{\includegraphics[width=7cm]{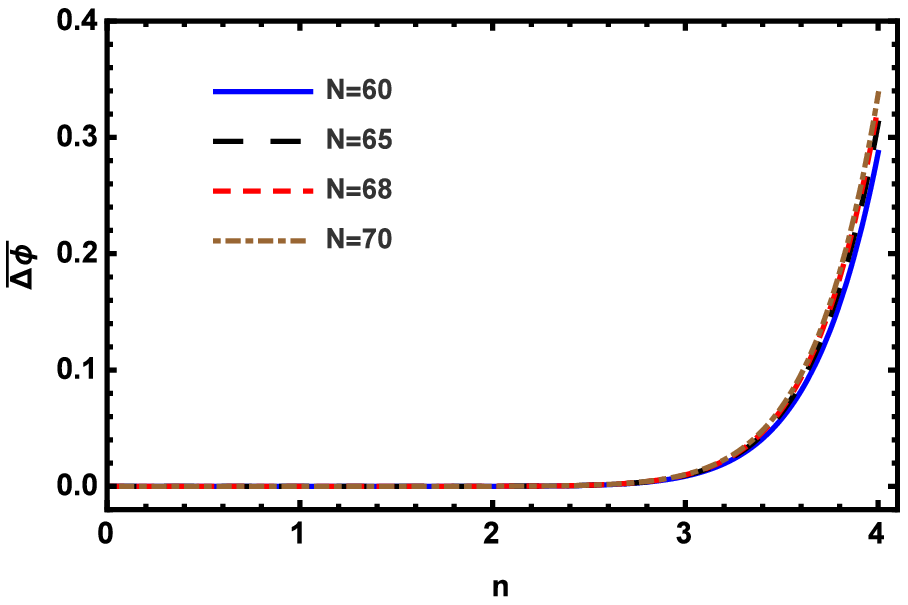}}
  \subfigure[\label{SC2A}]{\includegraphics[width=7.5cm]{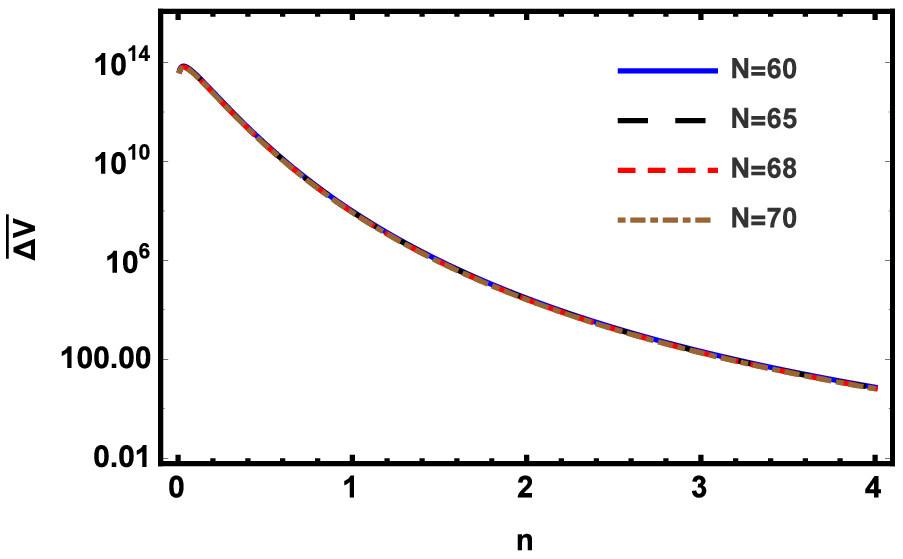}}
  \subfigure[\label{SC1B}]{\includegraphics[width=7cm]{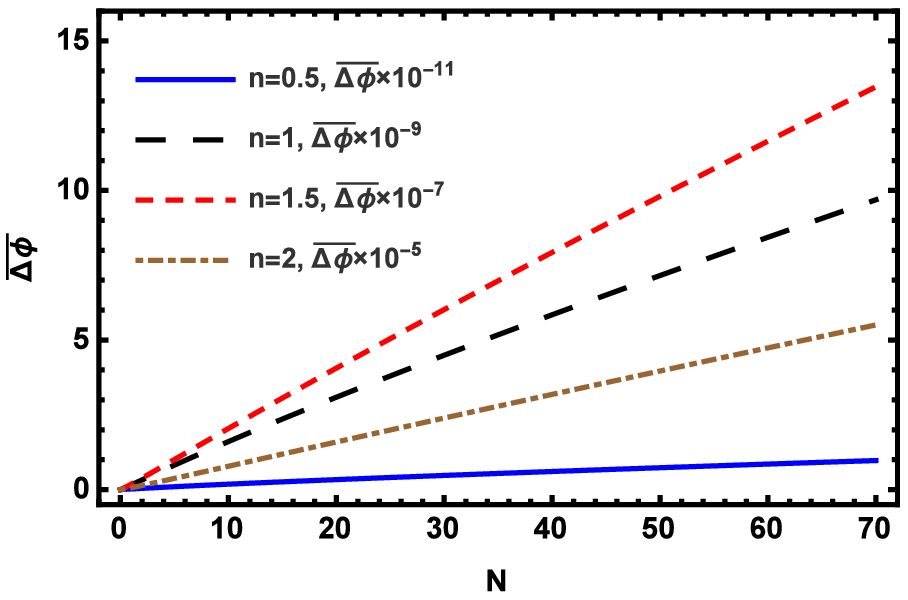}}
  \subfigure[\label{SC2B}]{\includegraphics[width=7.2cm]{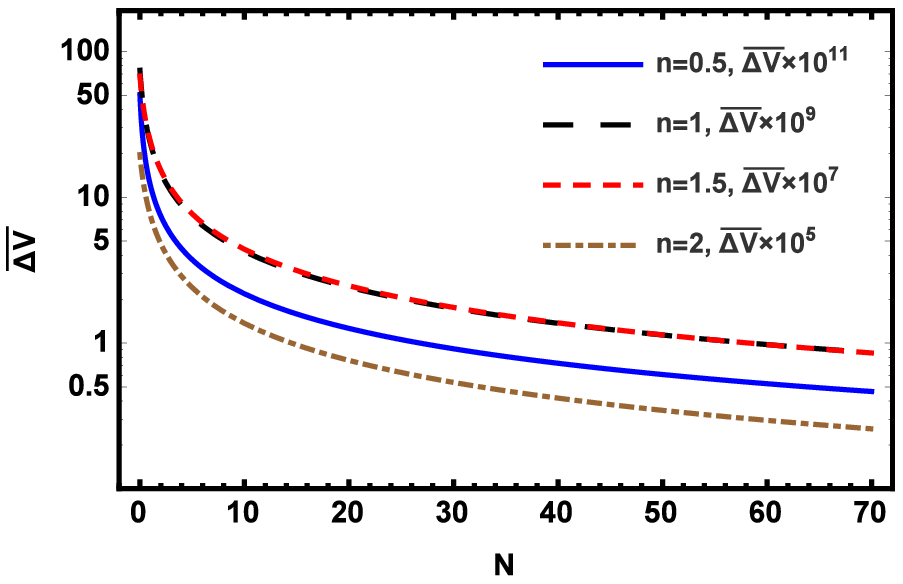}}
  \caption{The figures show the behavior of: a) $\Delta\phi/M_p \equiv \bar{\Delta\phi}$ and b) $M_p |V' / V| \equiv \bar{\Delta V}$ versus the number of e-fold for different values of $n$ taken from Fig.\ref{nN}; where $M_5=2 \times 10^{14} \; {\rm GeV}$.}\label{SCpower}
\end{figure}

Now that we determined the  model parameter space that are consistent with observation, the next step is to use these values and determine whether the model satisfy  the swampland criteria. For that,  in  Figs.\ref{SCpower}, we display the behavior of $\Delta\phi / M_p$ and $M_p |V' / V|$ for different values of the $n$ and the number of e-folds.  %Figs.\ref{SC1A} and \ref{SC2A} respectively  %represent  the behavior of $\Delta\phi / M_p$ and $M_p |V' / V|$ versus the parameter $n$ at the time of horizon crossing\footnote{Note  that different values of $N$ correspond  to  different horizon crossing %time.}.
Fig.\ref{SC1A} and  Fig.\ref{SC2A} respectively   show  that   $\Delta\phi / M_p < 1$   and  $M_p |V' / V| >1$ at the horizon crossing time for allowed values of $n$   (based on  Fig.\ref{nN}).
%  The behavior of $\Delta\phi / M_p$ and $M_p |V' / V|$ during the inflation are given in Figs.\ref{SC1B} and \ref{SC2B}, respectively, for different values of $n$ chosen from Fig.\ref{nN}.
From  Fig.\ref{SC1B}, we see that for all   the  chosen values of $n$, we have $\Delta\phi / M_p < 1$  during the whole time of inflation\footnote{By the "Whole time of inflation", the authors mean the time period between the horizon exit of perturbations and the end of inflation. The horizon exit is assumed to happen for number of inflation about $N=65$ and the end of inflation is illustrated by $N=0$.}, and it  gets even smaller as $n$ decreases.  On the other hand,  Fig.\ref{SC2B} shows that   $M_p |V'/ V| > 1$ for all  the chosen values of  $n$, and  gets even larger for  smaller $n$. We therefore conclude that for the all values of $n$ and $N$ presented in Fig.\ref{nN}, both swampland criteria are satisfied during the whole time of inflation. We also note that smaller values of the five dimensional Planck mass support the swampland criteria, namely by reducing the value of $M_5$, $\Delta\phi / M_p$  decreases and $M_p |V' / V|$ increases, respectively.

%On the other hand, the behavior of the term $M_p |V' / V|$ during inflation is displayed in Fig.\ref{SC2B}, where one could find that for all chosen values of $n$, the quantity is larger than one, and it is gets %even larger for the smaller values of $n$. Therefore, it could be concluded that for the all values of $n$ and $N$ presented in Fig.\ref{nN}, both swampland criteria are satisfied perfectly during the whole %time of inflation.} [\textcolor{green}{It seems to me from the figure this statement should be the other way around  $\Delta\phi / M_p$ is much larger and $M_p |V' / V|$ is much smaller; and also it seems this %is not always true, may be specify  for which n and the (range of ) N. Unless I miss understood your statement. Please try to re-state it}]}.
%for different value of $n$
% in  Figs.\ref{SC1} and \ref{SC2} respectively display the behavior of $\Delta\phi / M_p$ and $M_p |V / V'|$ during inflation for different value of $n$. It is seen that $\Delta\phi / M_p$ is much smaller than %unity and $M_p |V / V'|$ is much larger than unity which means that the model could perfectly satisfy the swampland criteria.
%%%%%%%%%%%%%%%%%%%%%%%%%%%%%%%%%%%%%%%%%%%%
%%%%%%%%%%%%%%%%%%%%%%%%%%%%%%%%%%%%%%%%%%%%
%We also note that the value of the five-dimensional Planck mass affects the results  in  which smaller values of $M_5$ support the swampland criteria, namely by reducing the value of $M_5$, $\Delta\phi / %M_p$  decreases and $M_p |V' / V|$ increases, respectively. However, for smaller $M_5$, the energy scale of inflation also decreases.
The same potential has been consider for the brane inflation in \cite{Lin:2018kjm} following a different method for considering the consistency of the model with observational data and swampland criteria. However, there are some differences which made it worth to reconsider the potential. Here, we were looking for the most suitable values of $n$ and $N$ which simultaneously give compatible results for the scalar spectral index and tensor-to-scalar ratio, in which for every values of the number of e-fold one could find the best choices of the parameter $n$ from Fig.\ref{nN} to have a good agreement with data. Besides, using the amplitude of the scalar perturbations, the constant $V_0$ has been determined as well which is used to find about the energy scale of the inflation in the model. To investigate the consistency of the model with the swampland criteria, both criteria have been considered in detail and the effect of $M_5$ on the criteria was studied as well. It stated that increasing of $M_5$ enhances the term $\Delta\phi / M_p$ in which at a certain value of $M_5$ the condition $\Delta\phi / M_p < 1$ is violated. The validity of two criteria were plotted for certain values of the parameters and during inflation which perfectly covers the subject.

%=============================================================%
%=============================================================%
%=============================================================%
%=============================================================%
%=============================================================%
\subsection{Natural Potential}
\begin{figure}[h]
  \centering
  \includegraphics[width=7cm]{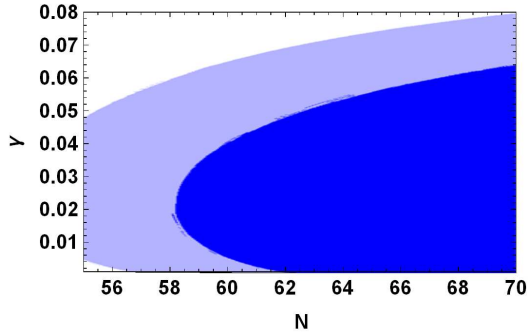}
  \caption{The  parameter space  $(\gamma, N)$  for the axion-like potential  that  yield  values $(r, n_s)$ allowed by  the Planck data at the $95\%$ CL  (light blue)and the $68\%$ CL (dark blue).
 }\label{NaturalgN}
\end{figure}

\begin{figure}[h]
  \centering
  \subfigure[\label{SCNatural01}]{\includegraphics[width=7.2cm]{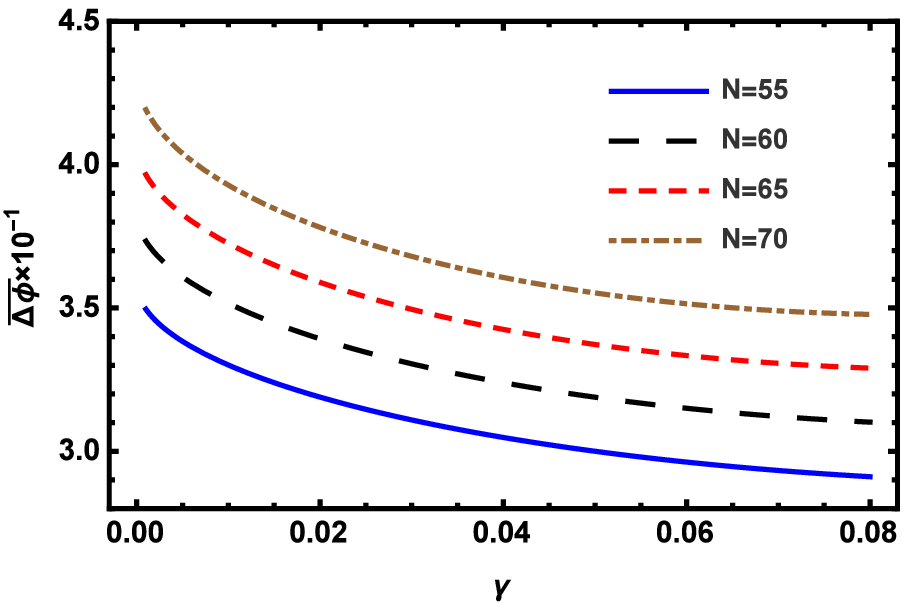}}
  \subfigure[\label{SCNatural02}]{\includegraphics[width=7cm]{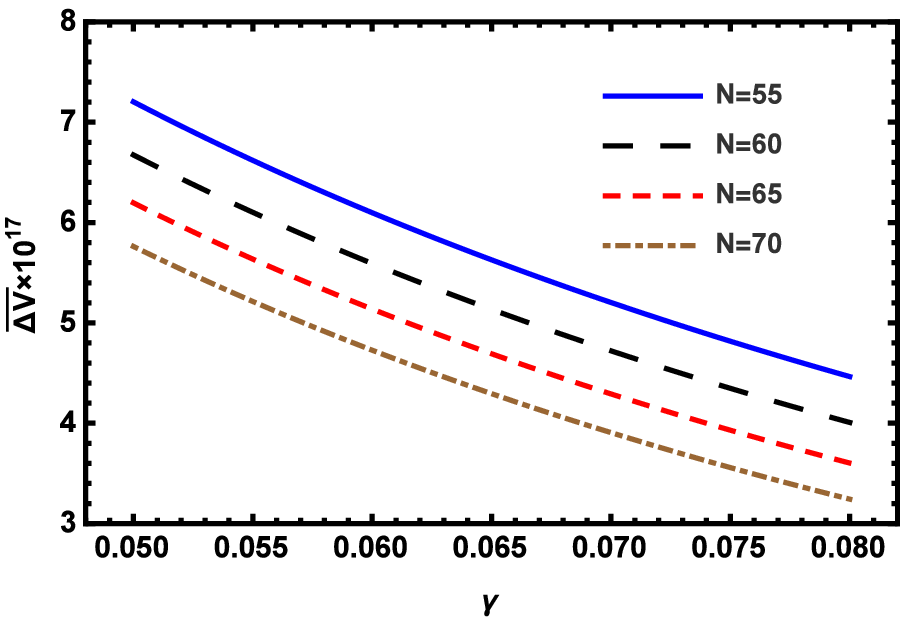}}
  \subfigure[\label{SCNatural03}]{\includegraphics[width=7.2cm]{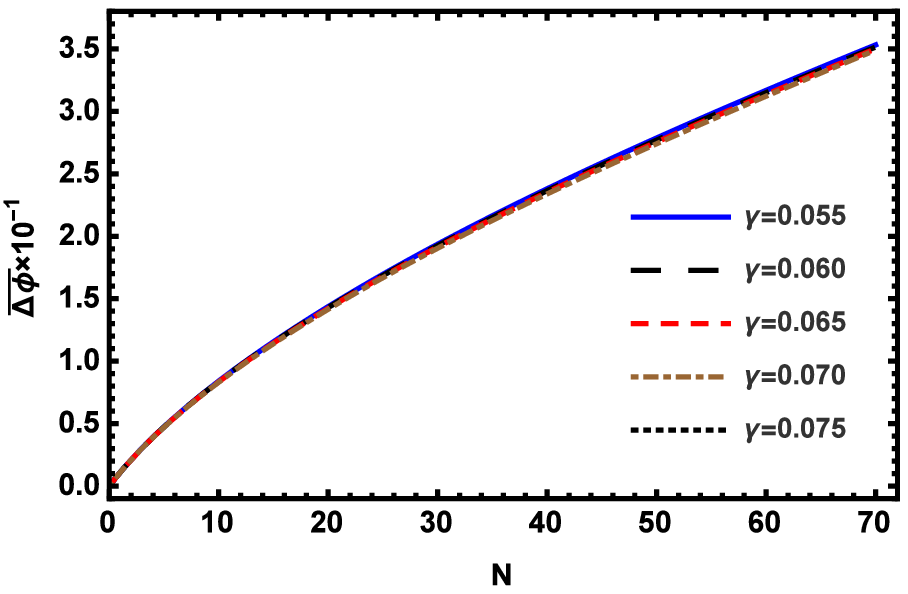}}
  \subfigure[\label{SCNatural04}]{\includegraphics[width=7cm]{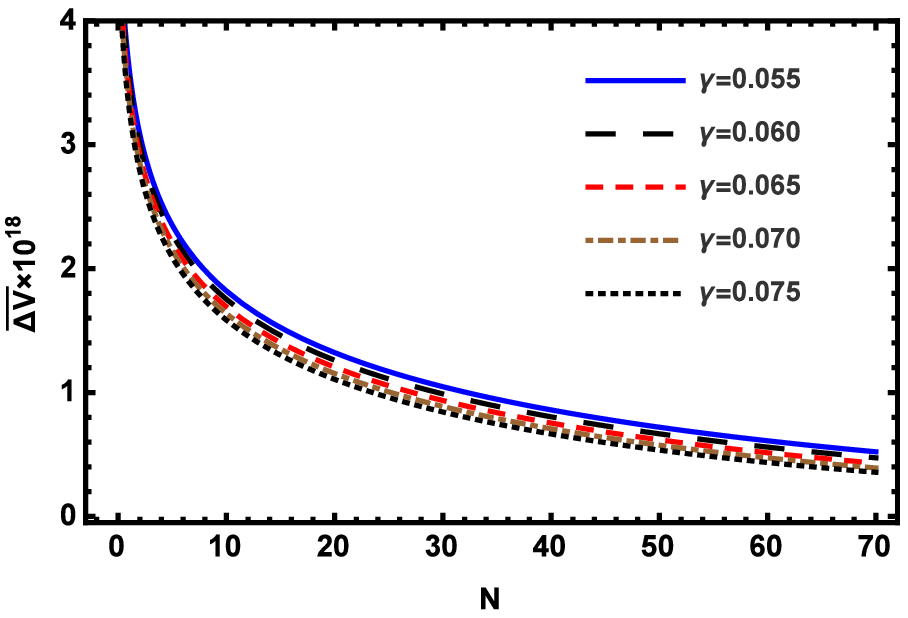}}
  \caption{The figures show the behavior of: a) $M_p \Delta\phi \equiv \bar{\Delta\phi}$ and b) $M_p V / V' \equiv \bar{\Delta V}$ versus the number of e-fold for different values of $n$ taken from Fig.\ref{NaturalgN}. The right panel of the figure displays the behavior of the axion-like potential.}\label{SCNatural}
\end{figure}

One of the interesting inflationary model is the natural inflation with the following potential
\begin{equation}\label{Vnatural}
  V(\phi) = V_0 \; \left( 1 - \cos\left( {\phi \over f} \right) \right),
\end{equation}
here   $V_0$ and $f$ are  constant parameters. This type of the potential usually arises when the inflaton is taken as an axion \cite{Baumann:2009ds}. The potential also has a strong background in particle physics model \cite{Adams:1992bn}. One of the features of the potential is that depending of the value of $f$ it could play both in large-field and small-field types inflation. For this potential, the slow-roll parameters  are given by %are obtained from Eq.\eqref{SRPV} as
\begin{equation}\label{SRP-Vnatural}
  \epsilon = \left( {3 M_5^3 \over 4\pi } \right)^2 \; {1 \over 3 f^2 V_0} \; {(1+\cos(\Phi)) \over (1-\cos(\Phi))^2}, \qquad
  \eta = \left( {3 M_5^3 \over 4\pi } \right)^2 \; {1 \over 3 f^2 V_0} \; {\cos(\Phi) \over (1-\cos(\Phi))^2}.
\end{equation}
where  $\Phi \equiv \phi / f$. At the end of inflation, we have
%where $\Phi \equiv \phi / f$. Reading the scalar field at the end of inflation from the relation $\epsilon = 1$, as
\begin{equation}\label{phie-Vnatural}
 \cos(\Phi_e) = {1 \over 2} \; \left[ (\gamma + 2) \pm \sqrt{(\gamma+2)^2 - 4 (\gamma-1)} \right], \qquad
\gamma \equiv \left( {3 M_5^3 \over 4\pi } \right)^2 \; {1 \over 3 f^2 V_0},
\end{equation}
and  after inserting it into Eq.\eqref{efold}, the field during  at horizon crossing reads
\begin{equation}\label{phis-Vnatural}
  \cos(\Phi_{\star}) = -1 - 2 W\left[ {-1 \over 2} \; \exp\left( {-1 \over 2} - \zeta \right) \right]
\end{equation}
where
\begin{equation*}
  \zeta \equiv = \gamma \; N + \cos(\Phi_e) - 2 \ln\left( 1 + \cos(\Phi_e) \right).
\end{equation*}
and $W[x]$ is the Lambert function. Substituting Eq.\eqref{phis-Vnatural} in Eq.\eqref{SRP-Vnatural}, one find  that  the scalar spectral index and tensor-to-scalar ratio  are only a function of the constant $\gamma$ and the number of e-folds $N$.  Then,   we  can extract the allowed values of the model  parameters $(\gamma,N)$  that yield values of $(r, n_s)$ in agreement with  Planck data, as shown in Fig.\ref{NaturalgN} .
%Then, using the Planck $r-n_s$ diagram, one could find the proper range for these parameters which for every point of $(\gamma,N)$ in this area, the model prediction comes to a perfect agreement with observational data. The area is illustrated in Fig.\ref{NaturalgN}.
%%%%%%%%%%%%%%%%%%%%%%%%%%%%%%%%%
%%%%%%%%%%%%%%%%%%%%%%%%%%%%%%%%%
On the other hand, after some algebra,  the amplitude of the scalar perturbation can be expressed as %which by substituting the potential and after some manipulation, one arrives at
\begin{equation}\label{Ps-natural}
 \mathcal{P}_s =  \left( {V_0 \over 75 \pi^2 \gamma^3 f^4}\right) \frac{\left( 1 - \cos(\Phi_s) \right)^5}{\left(1 + \cos(\Phi_s) \right)}%= \left( 75 \pi^2 \gamma^3 \mathcal{P}_s \right) \; {1 + \cos(\Phi_s) \over \left( 1 - \cos(\Phi_s) \right)^
\end{equation}
Using the observational data for $\mathcal{P}_s$,  the expression of the scalar field in Eq.\eqref{phis-Vnatural}, and the  values of $\gamma$ and $N$ from Fig.\ref{NaturalgN},  we   determine  the possible  values of the  other constants of the model as  presented in Table.\ref{TNatural}.
%%%%%%%%%%%%%%%%%%%%%%%
\begin{table}
  \centering
  \begin{tabular}{p{1.5cm}p{1cm}p{2.5cm}p{2.5cm}p{2.5cm}}
    \hline
     \quad $n$ &  $N$ & $f \; ({\rm GeV})$ &  $V_0 \; ({\rm GeV^4})$ & $V^\star \; ({\rm GeV^4})$ \\
    \hline
    $0.055$   & $55$ & $3.66 \times 10^{17} $ & $4.01 \times 10^{58} $   & $7.32 \times 10^{58} $ \\

    $0.060$   & $60$ & $3.72 \times 10^{17} $ & $3.56 \times 10^{58} $   & $6.68 \times 10^{58} $ \\

    $0.065$   & $65$ & $3.79 \times 10^{17} $ & $3.17 \times 10^{58} $   & $6.06 \times 10^{58} $ \\

    $0.070$   & $65$ & $3.73 \times 10^{17} $ & $3.04 \times 10^{58} $   & $5.84 \times 10^{58} $ \\

    $0.075$   & $70$ & $3.82 \times 10^{17} $ & $2.70 \times 10^{58} $   & $5.25 \times 10^{58} $ \\
    \hline
  \end{tabular}
  \caption{The constant $V_0$ and the energy scale of the inflation for different values of $(n,N)$ taken from Fig.\ref{NaturalgN} and $M_5=5 \times 10^{15} \; {\rm GeV}$}\label{TNatural}
\end{table}
%%%%%%%%%%%%%%%%%%%%%%%
%So far, a valid range for the parameters have been obtained, therefore the model with natural potential could properly describe inflation. The next step is to find out whether the model could satisfy the swampland criteria for the obtained values of the %parameter or not.
To see if the swampland criteria is met  in this type models, we  depict in  Fig.\ref{SCNatural}  the quantities  $\Delta\phi/M_p$ and $M_p|V' / V|$ for different values of $\gamma$ and the number of e-fold $N$.   For instance,  Figs.\ref{SCNatural01} and \ref{SCNatural02}
%we  present these parameters as function of  $\gamma$ for different values of the number of e-folds.These   two  later figures
 determine $\Delta\phi/M_p$ and $M_p|V' / V|$ at a specific time  during  inflation (horizon crossing time) for different values of $\gamma$. We note that  when $\gamma$ decreases,  both $\Delta\phi/M_p$ and $M_p|V' / V|$ decreases, however, $\Delta\phi/M_p$ remains smaller than unity and $M_p|V' / V|$ is still bigger than one.  On the other hand, Figs.\ref{SCNatural03} and \ref{SCNatural04}, display the behavior of these  quantities from the start  to the end for different values of $\gamma$, and as  inflation approaches the end,  $\Delta\phi/M_p$ decreases, while $M_p|V' / V|$ increases. Thus, in the  brane gravity   the axion-like  potential satisfy both swampland criteria.

%%%%%%%%%%%%%%%%%%%%%%%%%%%%%%%%%%%%%%%%%%%%
%%%%%%%%%%%%%%%%%%%%%%%%%%%%%%%%%%%%%%%%%%%%

%=============================================================%
%=============================================================%
%=============================================================%
%=============================================================%
%=============================================================%
\subsection{Exponential Potential}
\begin{figure}[h]
  \centering
  \includegraphics[width=6cm]{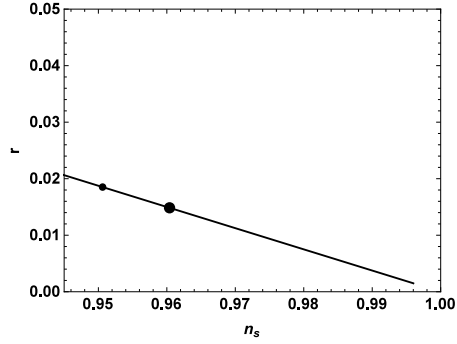}
  \caption{the figure shows the tensor-to-scalar ratio versus the scalar spectral index where the variable is the number of e-fold. The small and big point on the curve are respectively for $N=80$ and $N=100$.}\label{rnsN}
\end{figure}

One of the well-known potential in the inflationary studies is the exponential inflation which leads to a power-law inflation. The potential in given by
\begin{equation}\label{Vexponential}
  V(\phi) = V_0 \; \exp\left( \alpha \; \phi \right),
\end{equation}
where $V_0$ and $\alpha$ are two constants of the model. Substituting this potential in Eq.\eqref{SRPV}, the slow-roll parameters are found as
\begin{equation}\label{SRP-Vexp}
  \epsilon = \left( {3 M_5^3 \over 4\pi } \right)^2 \; {\alpha^2 \over 3 V_0} \; \exp\left( - \alpha \; \phi \right), \qquad
  \eta = \epsilon.
\end{equation}
Finding the scalar field at the end of inflation by solving the relation $\epsilon = 1$, and using that in Eq.\eqref{efold}, the scalar field during inflation in obtained in terms of the number of e-fold as
\begin{equation}\label{phis-Vexp}
  \exp\left( \alpha \; \phi_\star \right) = \left( {3 M_5^3 \over 4\pi } \right)^2 \; {\alpha^2 \over 3 V_0} \; (1+N)
\end{equation}
Then, the slow-roll parameters are given as
\begin{equation}\label{SRP-VexpN}
  \epsilon(N) = \eta(N) = \big( 1 + N \big)^{-1},
\end{equation}
and from Eqs.\eqref{ns} and \eqref{r}, the scalar spectral index and tensor-to-scalar ratio are obtained only as a function of the number of e-fold. Fig.\ref{rnsN} illustrates the behavior of the tensor-to-scalar versus the scalar spectral index in terms of the number of e-fold. The curve cross the region of $r-n_s$only for the number of e-fold $N>90$.
%%%%%%%%%%%%%%%%%%%%%%%%%%%%%%%%%
%%%%%%%%%%%%%%%%%%%%%%%%%%%%%%%%%

%=============================================================%
%=============================================================%
%=============================================================%
%=============================================================%
%=============================================================%
\subsection{T-mode Potential}
\begin{figure}[h]
  \centering
  \includegraphics[width=7cm]{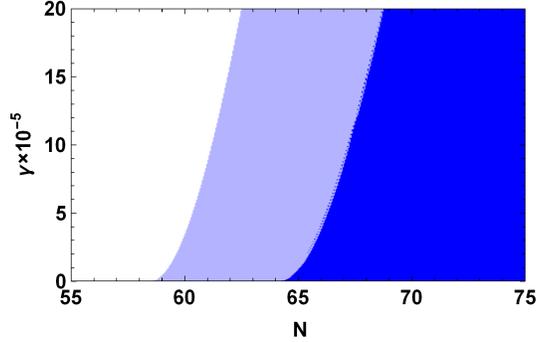}
  \caption{The allowed parameter space  $(\gamma, N)$  for the T-mode potential  that yield values of  $(r, n_s)$  that are in agreement with observation at the  $68\%$ Cl (dark blue)  and $95\%$ CL (light blue).}\label{TmodegN}
\end{figure}

The T-mode potential usually appears in the $\alpha$-attractor model of inflation which includes a non-canonical kinetic terms which is originated from Kahler potential in supergravity theories \cite{Kallosh:2013yoa,Kallosh:2013tua,Ferrara:2013rsa,Ferrara:2013kca,Dimopoulos:2017zvq}. This class of inflation includes Starobinsky’s inflation model and Higgs inflation model \cite{Ueno:2016dim}. The T-mode is one of the generalized models of the $\alpha$-attractors which is known with the following potential
\begin{equation}\label{TmodeV}
  V(\phi) = V_0 \tanh^2\left( { \phi \over \sqrt{6 \alpha} } \right).
\end{equation}
where $V_0$ and $\alpha$ are  free  constant parameters. The slow-roll parameters are  % \eqref{SRPV}, one arrives at
\begin{equation}\label{SRPTmode}
  \epsilon = {2\gamma \left( 1 - \tanh^2\left( { \phi \over \sqrt{6 \alpha} } \right) \right)^2 \over \tanh^4\left( { \phi \over \sqrt{6 \alpha} } \right)}, \qquad
  \eta = {\gamma \left( 1 - \tanh^2\left( { \phi \over \sqrt{6 \alpha} } \right) \right) \left( 1 - 3\tanh^2\left( { \phi \over \sqrt{6 \alpha} } \right) \right) \over \tanh^4\left( { \phi \over \sqrt{6 \alpha} } \right)},
\end{equation}
with  the  parameter $\gamma$ given by
\begin{equation*}
  \gamma \equiv \left( {3M_5^3 \over 4\pi} \right)^2 {1 \over 9 V_0 \alpha}
\end{equation*}
%After   finding  the scalar field at the end of inflation,  and substituting its expression  into the equation for the  the number of e-folds, we  can determine
The  scalar field at the horizon crossing time  is
\begin{equation}\label{phistarTmode}
  \cosh^2\left( { \phi_\star \over \sqrt{6 \alpha} } \right) - \ln\left( \cosh^2\left( { \phi_\star \over \sqrt{6 \alpha} } \right) \right) = \cosh^2\left( { \phi_e \over \sqrt{6 \alpha} } \right) - \ln\left( \cosh^2\left( { \phi_e \over \sqrt{6 \alpha} } \right) \right) + 2 \gamma N,
\end{equation}
Here $\phi_e$ is the value of the field at the end of inflation, given by %where
\begin{equation*}
  \cosh^2\left( { \phi_e \over \sqrt{6 \alpha} } \right) = 1 + \sqrt{2 \; \gamma}
\end{equation*}
%Therefore, from Eq.\eqref{phistarTmode}, it is realized that the slow-roll parameters at the horizon crossing time, and in turn the scalar spectral index and tensor-to-scalar, are expressed only in terms of the constant $\gamma$ and number of e-fold $N$.
Comparing the model predictions for  $n_s$ and $r$ with the Planck data, we  present in Fig.\ref{TmodegN}  the corresponding  allowed range of the constants  for $\gamma$ and $N$ at the  $68\%$ Cl (in dark blue)  and $95\%$ CL (in light blue).\\
%%%%%%%%%%%%%%%%%%%%%%%%%%%%%%%%%
%%%%%%%%%%%%%%%%%%%%%%%%%%%%%%%%%

Next,  the amplitude of scalar perturbations  at the crossing horizon time can be show to be  expressed as%just in terms of  the potential parameters as  %e  andin terms of the parameters  $\alpha$  and $\gamma$ %is related to $\gamma$ and $N$
\begin{equation}\label{V0Tmode}
  \alpha^3 = {1 \over (150\pi^2 \times 81) \; \gamma^4 \mathcal{P}_s} \; { \left( \cosh^2\left( { \phi_s \over \sqrt{6 \alpha} } \right) - 1 \right)^5 \over \cosh^6\left( { \phi_s \over \sqrt{6 \alpha} } \right)}
\end{equation}
Then, by choosing  specific values of $\gamma$ from the Fig.\ref{TmodegN}, the   allowed   potential parameters  $\alpha$ and $V_0$ can be determined and are shown  in Table.\ref{TTmode}. We also show in the values of  the  last two columns of the table  the values of $\Delta\phi/M_p$ and $M_p |V'/V|$ where we see that they  satisfy  the swampland criteria\footnote{Note that since the scalar field  approaches  $\phi_e$ as time passes, the quantity  $\Delta\phi/M_p$ gets smaller and smaller.}. Therefore,  the  T-mode potential  can be a viable  model for inflation that satisfies the swampland criteria.  %%%%%%%%%%%%%%%%%%%%%%%
\begin{table}
  \centering
  {\footnotesize
  \begin{tabular}{p{2cm}p{1cm}p{2.3cm}p{2.3cm}p{2.3cm}p{1.5cm}p{1.5cm}}
    \hline
     $\qquad \gamma$ &  $N$ &  $\alpha \; ({\rm GeV^{2}})$ &  $V_0 \; ({\rm GeV^4})$ &  $V^\star \; ({\rm GeV^4})$ & ${\Delta\phi/M_p}$ & $M_p |V'/V|$\\
    \hline
    $1.5 \times 10^{-5}$   & $66$ & $4.01 \times 10^{36} $     & $1.64 \times 10^{60} $
    & $9.95 \times 10^{58} $     & $0.406$  & \quad $3.33$  \\

    $3 \times 10^{-5}$ & $67$ & $2.84 \times 10^{36} $ & $1.15 \times 10^{60} $
    & $9.83 \times 10^{58} $     & $0.409$  & \quad $3.25$  \\

    $5 \times 10^{-5}$   & $69$ & $3.65 \times 10^{36} $   & $8.86 \times 10^{59} $
    & $9.68 \times 10^{58} $     & $0.416$ & \quad $3.15$  \\

    $6.5 \times 10^{-5}$ & $71$ & $1.98 \times 10^{36} $ & $7.65 \times 10^{59} $
    & $9.55 \times 10^{58} $     & $0.423$  & \quad $3.07$  \\

    $8 \times 10^{-5}$   & $73$ & $1.82 \times 10^{36} $   & $6.78 \times 10^{59} $
    & $9.42 \times 10^{58} $     & $0.430$   & \quad $2.49$  \\
    \hline
  \end{tabular}
  }
  \caption{The constants $\alpha$, $V_0$ and the energy scale of the inflation for different values of $(\gamma,N)$ taken from Fig.\ref{TmodegN} and $M_5=5 \times 10^{15} \; {\rm GeV}$. Also, the last two columns of the table determine the $\Delta\phi/M_p$ and $M_p |V'/V|$ and give some insight about the swampland criteria.}\label{TTmode}
\end{table}
%%%%%%%%%%%%%%%%%%%%%%%
%Finally, we are going to examine the swampland criteria for this these values.
%An inspection of the last two columns of the table, show  that $\Delta\phi/M_p$  is  smaller than unity
 %the values of $\Delta\phi/M_p$ and $M_p |V'/V|$. The term $\Delta\phi/M_p$ could be smaller than unity, and since by passing time the scalar field approaches to $\phi_e$ it could be concluded that the term $\Delta\phi/M_p$ gets smaller and smaller. %Therefore, the first criteria $\Delta\phi/M_p < c$ perfectly is satisfied. The last column determines that the term
%$M_p |V'/V|$ is larger than unity and the second criterion $M_p |V'/V| > c'$ will be satisfied as well. Then, the model for the T-mode potential could be perfectly match with the observational data, and on the other hand it is able to appropriately satisfy the two %swampland criteria.\\

%=============================================================%
%=============================================================%
%=============================================================%
%=============================================================%
%=============================================================%
\subsection{Generalized T-mode Potential}
\begin{figure}[h]
  \centering
  \includegraphics[width=7cm]{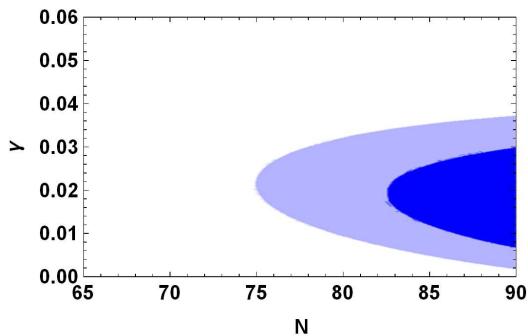}
  \caption{The allowed parameter space  $(\gamma, N)$  for the generalized T-mode potential  that yield values of  $(r, n_s)$  that are in agreement with observation at the  $68\%$ Cl (dark blue)  and $95\%$ CL (light blue).}\label{gNGT}
\end{figure}

Here we consider a slightly modified T-mode potential
\begin{equation}\label{GTmodeV}
  V(\phi) = V_0 \left( 1 - \tanh^2\left( \alpha \phi \right) \right)
\end{equation}
%where $V_0$ and $\alpha$ are two constants of the model that are going to be determined later. Inserting this potential in Eq.\eqref{SRPV},
Following similar steps as  we did with the previous type of potentials, we obtain  the slow-roll parameters at the crossing time %are obtained as
%\begin{equation}\label{SRPGT}
%  \epsilon = {\gamma \tanh^2\left( \alpha \phi \right) \over 1 - \tanh^2\left( \alpha \phi \right) }, \qquad
%  \eta = {-\gamma \over 2} \; {1 - 3\tanh^2\left( \alpha \phi \right) \over 1 - \tanh^2\left( \alpha \phi \right)}
%\end{equation}
%where the defined constant $\gamma$ here is given by
%\begin{equation*}
%  \gamma \equiv \left( 3M_5^3 \over 4\pi \right)^2 {4\alpha^2 \over 3V_0}.
%\end{equation*}
%Extracting the scalar field at the end of inflation, i.e. by solving $\epsilon(\phi_e) = 1$ and by using the result in Eq.\eqref{efold}, the scalar field at horizon crossing time is obtained as
%\begin{equation}\label{phistarGT}
%  \tanh^2\left( \alpha \phi_\star \right) = \tanh^2\left( \alpha \phi_e \right) \; \exp\left( - \gamma N \right), \qquad
%  \tanh^2\left( \alpha \phi_e \right) = {1 \over 1+\gamma}
%\end{equation}
%Then, substituting Eq.\eqref{phistarGT} in \eqref{SRPGT}, one arrives at
\begin{equation}\label{SRPstarGT}
  \epsilon = { \gamma \; \exp\left( - \gamma N \right) \over (1+\gamma) - \exp\left( - \gamma N \right)}, \qquad
  \eta = {-\gamma \over 2} \; {(1+\gamma) - 3 \exp\left( - \gamma N \right) \over (1+\gamma) - \exp\left( - \gamma N \right)}
\end{equation}
where the defined constant $\gamma$ here is given by
\begin{equation*}
  \gamma \equiv \left( 3M_5^3 \over 4\pi \right)^2 {4\alpha^2 \over 3V_0}.
\end{equation*}
%Same as previous cases, it is realized from Eqs.\eqref{ns} and \eqref{r} that the scalar spectral index and tensor-to-scalar ratio are obtained only as a function of the constant $\gamma$ and the number of %e-fold $N$.
By  comparing the model predictions for $n_s$ and $r$ with the Planck $r-n_s$ diagram, we find that   only for small  range  of the parameters $\gamma$ and $N$ the model is in   agreement with  the observational data, as depicted in  Fig.\ref{gNGT}
%%%%%%%%%%%%%%%%%%%%%%%%%%%%%%%%%
%%%%%%%%%%%%%%%%%%%%%%%%%%%%%%%%%

For the amplitude of the scalar perturbations at the horizon crossing time, we find % it is found that
\begin{equation}\label{GTV0}
  V_0 \alpha^4 = {75 \pi^2 \mathcal{P}_s \over 16 \gamma^3} \; { (1+\gamma)^3 \exp\left( -\gamma N \right) \over \left( (1+\gamma) - \exp\left( -\gamma N \right) \right)^4}
\end{equation}
Thus, for a given  point  of   the allowed region in  Fig.\ref{gNGT}, we can   use of  the data for the amplitude of the scalar perturbations to   determine the  parameters of the potential. In  Table.\ref{TGT}  we  give   the values of the constants $\alpha$ and $V_0$  for a chosen set  of points  from Fig.\ref{gNGT}.
%%%%%%%%%%%%%%%%%%%%%%%
\begin{table}
  \centering
  {\footnotesize
  \begin{tabular}{p{1.5cm}p{1cm}p{2.5cm}p{2.5cm}p{2.5cm}p{2.5cm}p{2.5cm}}
    \hline
     $\quad \gamma$ &  $N$ &  $\alpha \; ({\rm GeV^{-1}})$ &  $V_0 \; ({\rm GeV^4})$ &  $V^\star \; ({\rm GeV^4})$ \\
    \hline
    $0.010$   & $89$ & $2.94 \times 10^{-16} $     & $4.21 \times 10^{61} $
    & $2.50 \times 10^{61} $      \\

    $0.015$   & $84$ & $2.12 \times 10^{-16} $ & $1.46 \times 10^{61} $
    & $1.05 \times 10^{61} $    \\

    $0.020$   & $76$ & $1.74 \times 10^{-16} $   & $7.38 \times 10^{60} $
    & $7.38 \times 10^{60} $    \\

    $0.025$   & $80$ & $1.39 \times 10^{-16} $ & $3.79 \times 10^{60} $
    & $3.29 \times 10^{60} $     \\

    $0.030$   & $85$ & $1.15 \times 10^{-16} $   & $2.13 \times 10^{60} $
    & $1.97 \times 10^{60} $     \\
    \hline
  \end{tabular}
  }
  \caption{The constants $\alpha$, $V_0$ and the energy scale of the inflation for different values of $(\gamma,N)$ taken from Fig.\ref{gNGT} and $M_5=2 \times 10^{14} \; {\rm GeV}$}\label{TGT}
\end{table}
%%%%%%%%%%%%%%%%%%%%%%%

%The model constants have been determined properly in which for them the model comes to a good consistency with observational data. Now, we are going to examine whether for these constants the model could also satisfy the swampland criteria.

\begin{figure}
  \centering
  \subfigure[\label{SCGTmode01}]{\includegraphics[width=7cm]{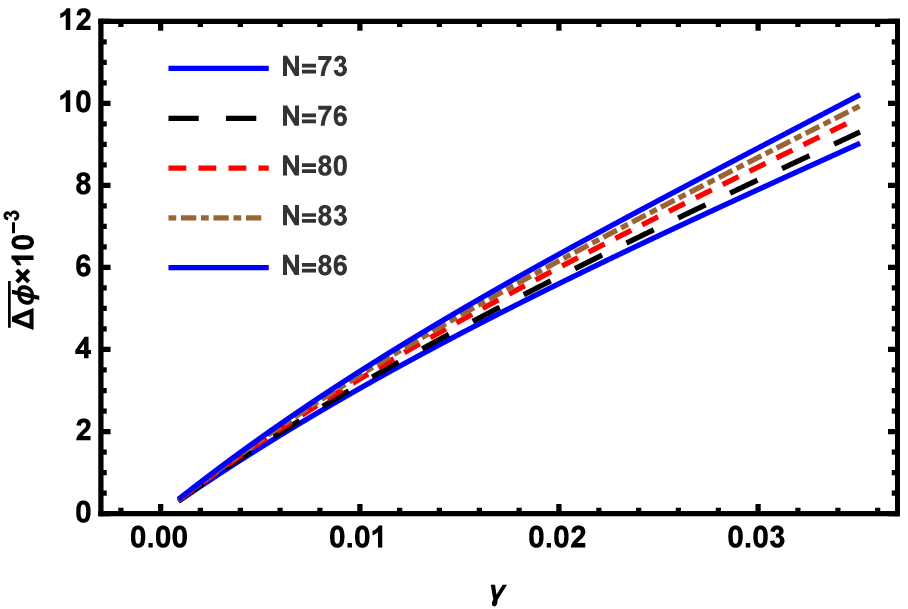}}
  \subfigure[\label{SCGTmode02}]{\includegraphics[width=7.2cm]{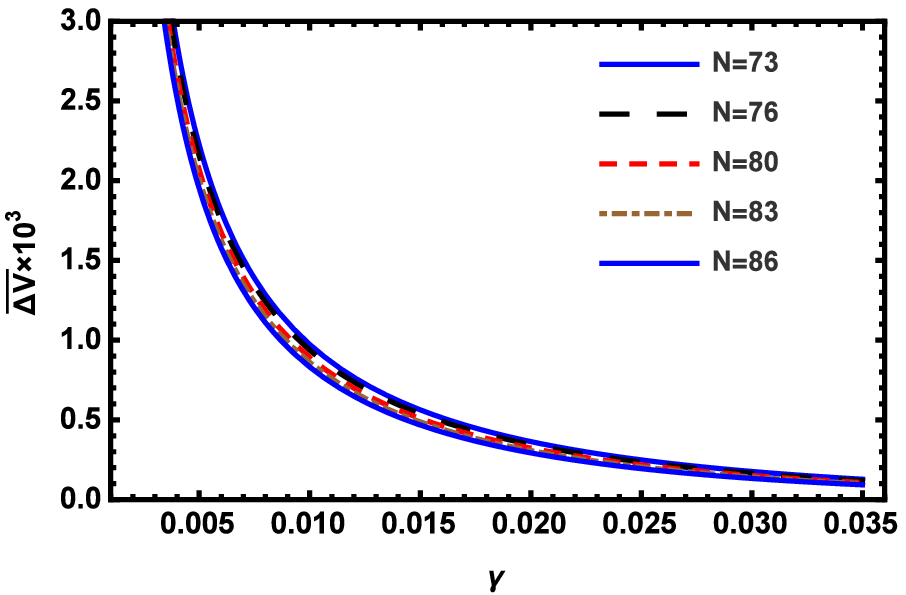}}
  \subfigure[\label{SCGTmode03}]{\includegraphics[width=7.1cm]{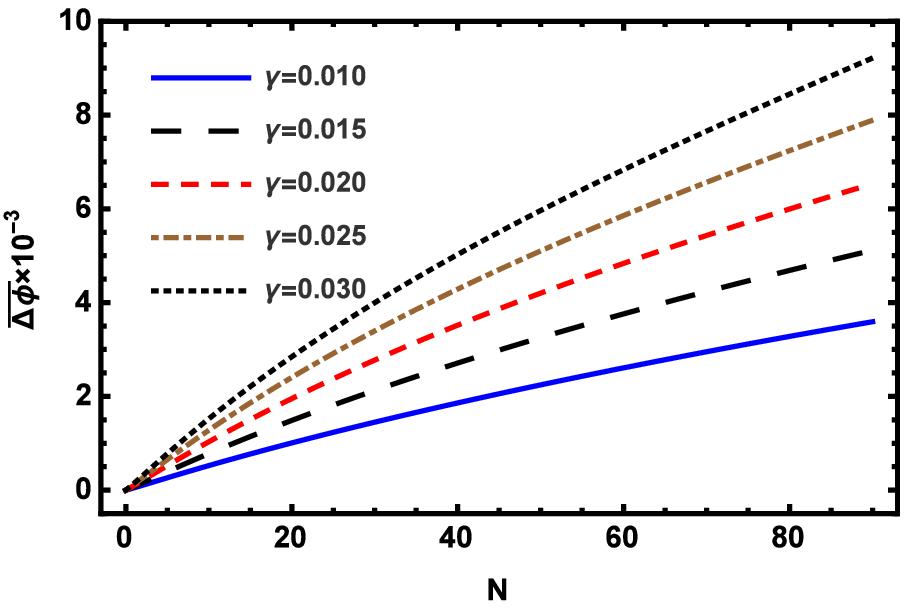}}
  \subfigure[\label{SCGTmode04}]{\includegraphics[width=7cm]{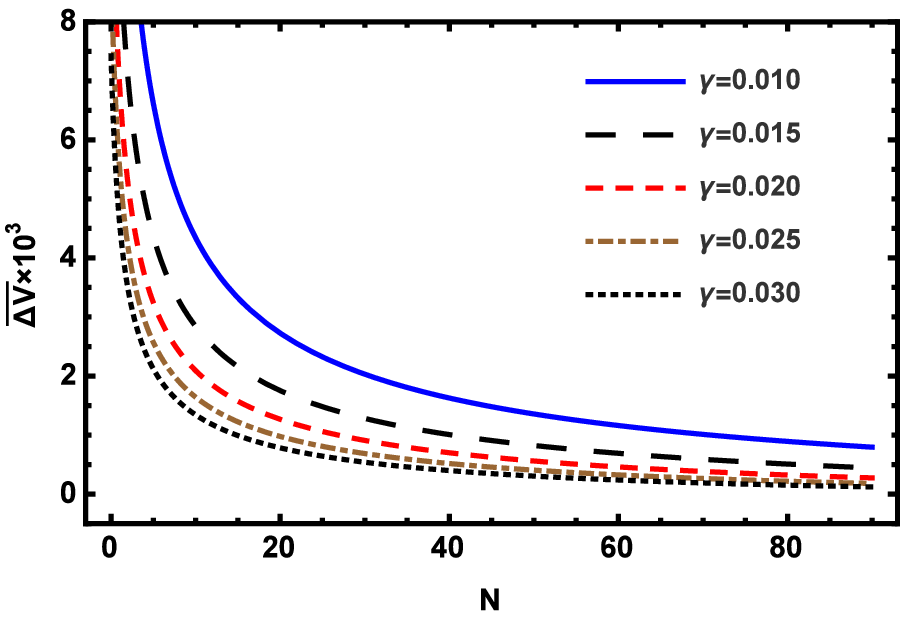}}
  \caption{The top row shows the behavior of $M_p \Delta\phi \equiv \bar{\Delta\phi}$ and $M_p V / V' \equiv \bar{\Delta V}$ versus $\gamma$ for different values of number of e-fold. The bottom row indicates the behavior of $M_p \Delta\phi \equiv \bar{\Delta\phi}$ and $M_p V / V' \equiv \bar{\Delta V}$ versus $N$ for different values of $\gamma$ taken from Fig.\ref{gNGT}.}\label{SCGTmode}
\end{figure}

To examine the swampland criteria,  in Fig.\ref{SCGTmode} we  plot  the  quantities $\Delta\phi/M_p$ and $M_p |V' / V|$ for different values of $\gamma$ and $N$. %In Figs.\ref{SCGTmode01} and %\ref{SCGTmode02} the terms are plotted versus $\gamma$ for different values of the number of e-fold.
%The Plots actually express the terms in a specific time (indicates by the number of e-fold) for different values of $\gamma$ obtained in Fig.\ref{gNGT}.
In Figs.\ref{SCGTmode01} and \ref{SCGTmode02} we  note that as $\gamma$ increases, $\Delta\phi/M_p$ and $M_p |V' / V|$ respectively increases and decreases. Figs.\ref{SCGTmode03} and \ref{SCGTmode04}  which represent  the  behavior  of these quantities    during inflation (versus the number of e-fold) for different values of the constant $\gamma$, and  as the inflaton      approaches  the end of inflation, $\Delta\phi/M_p$ and $M_p |V' / V|$ respectively decreases (as was expected) and increases, and hence  during the whole period  of inflation the swampland criteria are satisfied.  Therefore, the potential of the form \eqref{GTmodeV} can be in consistent with the Planck data and satisfy the swampland criteria.

%%%%%%%%%%%%%%%%%%%%%%%%%%%%%%%%%%%%%%%%%%%%
%%%%%%%%%%%%%%%%%%%%%%%%%%%%%%%%%%%%%%%%%%%%

%=============================================================%
%=============================================================%
%=============================================================%
%=============================================================%
%=======================  Section 5 ==========================%
%=============================================================%
%=============================================================%
%=============================================================%
%=============================================================%
\section{Reheating}\label{reheating}
After inflation the universe is cold and almost empty of particles. Then, a mechanism is required to heat up the universe and allow for a smooth cross to radiation dominant phase. The mechanism is called (p-)reheating. The physics of reheating is still uncertain but in a simple words, it could be said that the energy stored in scalar field is converted to the relativistic particle, so that after this period the universe ends up with the standard radiation dominant era.  \\
During inflation, the universe expriences an expansion about $N_k$ e-fold, and in the rehating phase it expands for another $N_{re}$ e-fold. The universe fluid is described by an effective equation of state (EoS), parametrize by $\omega_{eos}$, in which it is close to $-1$ during inflation, then it comes to $-2/3$ at the end of inflaiton\footnote{for this values of $\omega_{eos}$ the acceleration of the universe vanishes. Note that in the standard 4D cosmology, it happends for $\omega_{eos}=-1/3$.}. \\
In the reheating phase, the universe heats up and reaches to the final temperature $T_{re}$, which should stand in the range $10^{-3} \; {\rm GeV} < T_{re} <  10^{9-10}\; {\rm GeV}$ \cite{Kofman:1994rk,Kofman:1997yn,Bassett:2005xm,Allahverdi:2010xz,Amin:2014eta}. The well-known radiation dominant phase is initiated at this temperature $T_{re}$. The reheating number of e-fold and temperature could be used as another way for constraining the model parameters. \\
The number of e-fold during the reheating phase is given by
\begin{equation}\label{Nr}
  N_{re} = \ln\left( {a_{re} \over a_{e}} \right).
\end{equation}
where $a_e$ and $a_{re}$ are respectively the scalar field at the end of inflation and the end of reheating. \\ 
One way to study the reheatin phase (followed by many such as \cite{Liddle:2003as,Dai:2014jja,Cook:2015vqa,Bhattacharjee:2016ohe,Drewes:2017fmn}) is to assume that the universe is dominated by an energy density which is conserved and is parametrized by an effective equation of state parameter, $\omega_{eos}$. Following this approach, the energy density at the end of inflation $\rho_{e}$ is related to the energy density at the end of reheating $\rho_{re}$ through the effective equation of state parameter as
\begin{equation}\label{rhor}
  {\rho_{e} \over \rho_{re}} = \left( {a_{e} \over a_r} \right)^{-3(1+\omega_{eos})}
\end{equation}
where $\omega_{eos}$ is assumed to be constant. The energy density at the end of inflation (where $\epsilon = 1$) is obtained as $\rho_e = (1 + \lambda) \; V_e$, where the parameter $\lambda = \left( {3 \over \epsilon} - 1 \right)^{-1}$ is the ratio of kinetic energy to the potential of the inflaton \cite{Dai:2014jja}. Also, the energy density at the end of reheating, which is in the form of radiation, is expressed in terms of the reheating temperature $T_{re}$ as $\rho_{re} = \big( \pi^2 g_{re} / 30 \big) \; T_r^4$ \cite{Liddle:2003as,Dai:2014jja,Cook:2015vqa,Bhattacharjee:2016ohe,Drewes:2017fmn}. 
%\begin{equation}\label{rhoT}
%  \rho_{re} = {\pi^2 g_{re} \over 30} \; T_r^4
%\end{equation}
The scenario of reheating still contains many uncertainties and there are many complicated factor to be included. According to the above approach, the reheating is modeled by one effective equation of state parameter $\omega_{eos}$. In the canonical reheating scenario, $\omega_{eos} = 0$, and a model with simple potential $V \propto \phi^2$ is also consistent with $\omega_{eos} = 0$ \cite{Abbott:1982hn,Albrecht:1982mp,Dolgov:1982th}. Models with $V \propto \phi$ and $\phi^{2/3}$ are possible to be more compatible with $-1/3 < \omega_{eos} < 0$ \cite{McAllister:2008hb,Silverstein:2008sg}. One the other hand, numerical studies gives the range $0 \lesssim \omega_{eos} \lesssim 0.25$ \cite{Podolsky:2005bw}. However, the main point is that the effective equation of state parameter must be $\omega_{eos} > -1/3$ to put an end on inflation (in brane cosmology, it is $\omega_{eos} > -2/3$). Here, we try to present some suggestions about the value of the parameter based on our finding in the phase of inflation.  \\

After some manipulation, the reheating e-fold and temperature are extracted as \cite{Martin:2010kz,Liddle:2003as,Dai:2014jja,Cook:2015vqa,Bhattacharjee:2016ohe,Drewes:2017fmn,Mohammadi:2020twg}
\begin{eqnarray}
  N_r &=& {-4 \over (1-3\omega)} \; \left[ {1 \over 4} \; \ln\left( {30 \over \pi^2 g_{re}} \right) +
                                           {1 \over 3} \; \ln\left( {11 g_{re} \over 43} \right)  + \ln\left( {k \over a_0 T_0} \right) + \ln\left( {\rho_{end} \over H_k} \right)  + N_k \right] \\
  T_r &=& \left( {43 \over 11 g_{re}} \right)^{1 / 3} \; {a_0 T_0 \over k} \; H_k \; e^{-N_k} \; e^{N_r}
\end{eqnarray}

In general there are a lower bound and an upper bound on the magnitude of reheating temperature. After inflation, the energy stored in inflaton decays to other particles, and they interact to reach to a thermal equilibrium. In order to attach to the hot big bang theory and have a successful big bang nucleosynthesis, the temperature of the thermal equilibrium should be greater than $1 \; {\rm MeV}$ \cite{Lozanov:2019jxc,Allahverdi:2000ss}. On the other side, to avoid the overproduction of gravitinos (unwanted particle) the temperature should be lower than $10^{9-10} \; {\rm GeV}$ \cite{Lozanov:2019jxc,Allahverdi:2000ss}. Consequently, it is expected that the reheating temperature stands in the range $1 \; {\rm MeV} < T_r < 10^{9-10} \; {\rm GeV}$. \\
The behavior of the reheating number of e-fold and temperature for all cases have been illustrated in Fig.\ref{TNreheatingPower} to \ref{TNreheatingGTmode} as follows. \\
\begin{itemize}
  \item \textbf{Power-law potential case:} The behavior of $N_r$ and $T_r$ versus the equation of the state parameter $\omega$ has been illustrated in Fig.\ref{TNreheatingPower} for different values of $n$ and $N$ taken from parametric space Fig.\ref{nN}. Since, the universe is expanding, only the positive values of $N$ is acceptable. Then, the parameter $\omega$ should stand between $-1/3 < \omega < 1/3$. Also, since the number of e-fold is expected to be of the order of one, the parameter should be chosen close to $-1/3$.
      %%%%%%%%%%%%%%%%%%%%%%%%%%%%%%%%%%%%%%%%%%%%%%%%%%%%%%%%%%
      \begin{figure}[h]
        \centering
        \subfigure[]{\includegraphics[width=7.5cm]{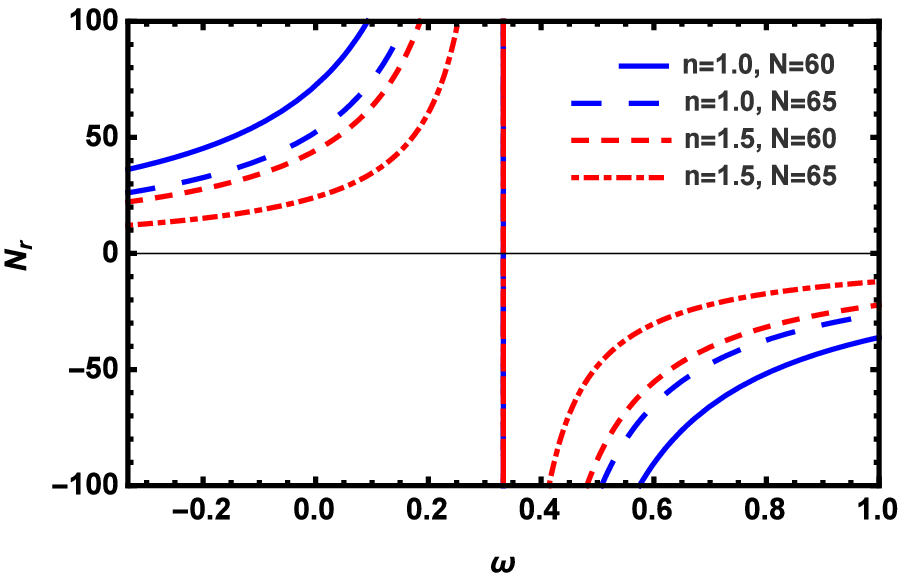}}
        \subfigure[]{\includegraphics[width=7cm]{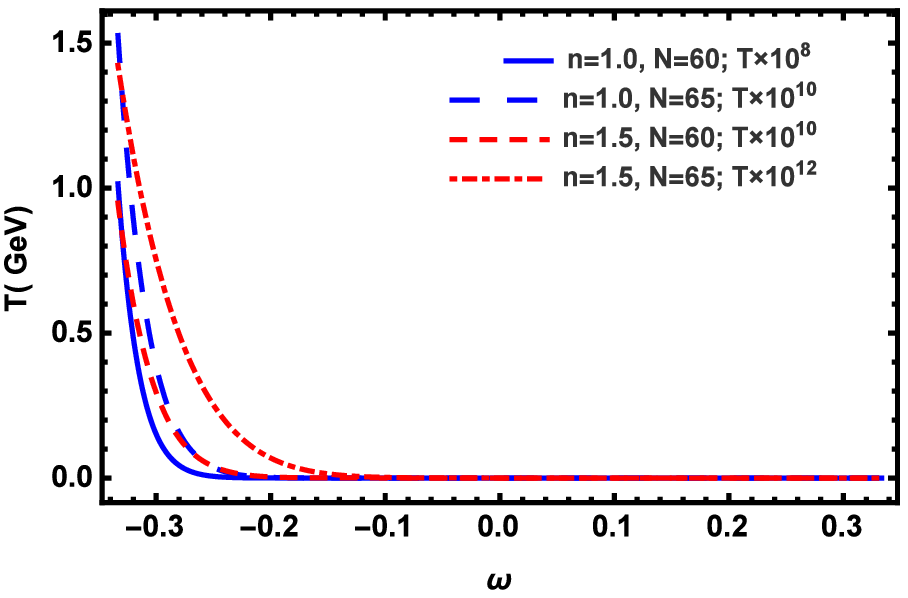}}
        \caption{The behavior of $N_r$ and $T_r$ versus the equation of the state parameter $\omega$ for different values of $n$ and $N$ taken from parametric space Fig.\ref{nN}. }\label{TNreheatingPower}
      \end{figure}
      %%%%%%%%%%%%%%%%%%%%%%%%%%%%%%%%%%%%%%%%%%%%%%%%%%%%%%%%%%%
      The temperature, on the other hand, is plotted for $-1/3 < \omega < 1/3$ (where the number of e-fold is positive), which indicates that when $\omega$ is close to $-1/3$ the temperature depends on the values of $n$ and number of e-fold. By increasing these parameters the temperature enhances as well. \\
      The reheating temperature depends on both $n$ and $N$ as clear from Fig.\ref{TNreheatingPower}. The model could produces the reheating temperature in the mentioned range, however, for by increasing both $n$ and $N$ it increase and in some point it will be larger than $10^{10} \; {\rm GeV}$. for example for $n=1.5$ and $N=65$, the temperature is about $T_r \sim 10^{12}$ which is above the upper bound. Therefore, another constraint could be imposed on the free parameters of the model and limit the parametric space of Fig.\ref{nN}.
  \item \textbf{Natural potential case:} Fig.\ref{TNreheatingNatural} presents the behavior of $N_r$ and $T_r$ versus the parameter $\omega$ for different values of $\gamma$ and $N$ taken from parametric space Fig.\ref{NaturalgN}. For this case, the number of e-fold is positive only for $1/3 < \omega < 1$, and the more interested values of $N_r$ occurs when $\omega$ is close to $1$. \\
      %%%%%%%%%%%%%%%%%%%%%%%%%%%%%%%%%%%%%%%%%%%%%%%%%%%%%%%%%%
      \begin{figure}[h]
        \centering
        \subfigure[]{\includegraphics[width=7.5cm]{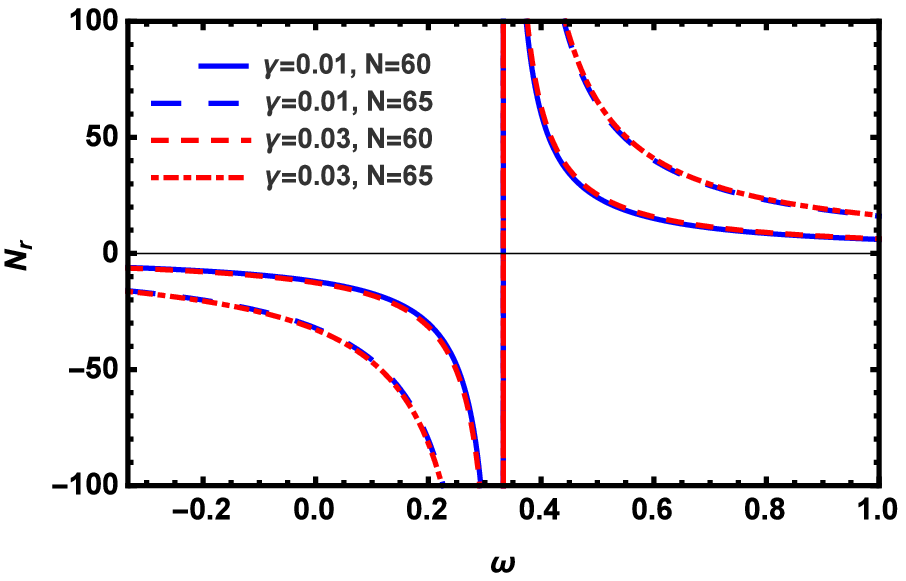}}
        \subfigure[]{\includegraphics[width=7cm]{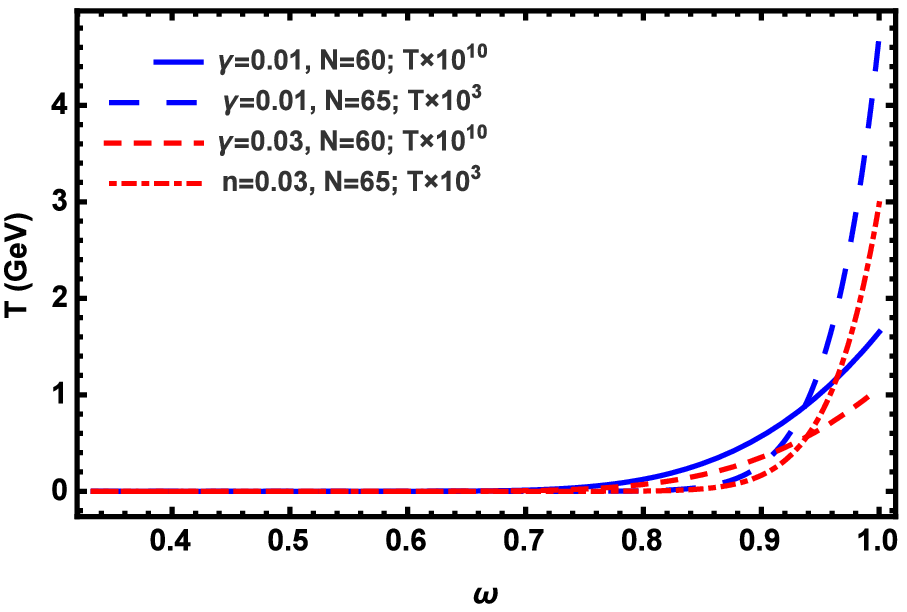}}
        \caption{The behavior of $N_r$ and $T_r$ versus the equation of the state parameter $\omega$ for different values of $\gamma$ and $N$ taken from parametric space Fig.\ref{NaturalgN}. }\label{TNreheatingNatural}
      \end{figure}
      %%%%%%%%%%%%%%%%%%%%%%%%%%%%%%%%%%%%%%%%%%%%%%%%%%%%%%%%%%%
      The temperature is depicted for $1/3 < \omega < 1$ (where $N_r$ is positive) for different values of $\gamma$ and $N$. It is realized that the temperature is more sensitive to the values of number of e-fold than the value of $\gamma$. For instance for $N=65$, the reheating temperature is of order of $T_r \propto 10^{3} \; {\rm GeV}$ and for $N=65$ it is
      about $T_r \propto 10^{10} \; {\rm GeV}$. \\
      The temperature $T_r$ is more sensitive to the number of e-fold $N$ and it increases by reduction of $N$. The reheating temperature cross the upper bound for some specific value of $N$, for example for $\gamma = 0.01$ and $N=60$ we have $T_r \sim 10^{10} \; {\rm GeV}$. Therefore, the acceptable range of reheating temperature applies a lower limit for the number of e-fold.
  \item \textbf{T-mode potential case:} For the T-mode potential, the behavior of the reheating number of e-fold and temperature is plotted in Fig.\ref{TNreheatingTmode} for the different values of the constant $\gamma$ and the number of e-fold $N$, taken from Fig.\ref{TmodegN}. The reheating number of e-fold is negative for the range $-1/3 < \omega < 1/3$ which is not acceptable since the universe is still expanding in the reheating phase. On the other hand, it is not desirable to have a large number of e-fold during reheating. Therefore, only the values of $\omega$ that are close to $1$ are more of interest. \\
      %%%%%%%%%%%%%%%%%%%%%%%%%%%%%%%%%%%%%%%%%%%%%%%%%%%%%%%%%%
      \begin{figure}[h]
        \centering
        \subfigure[]{\includegraphics[width=7.5cm]{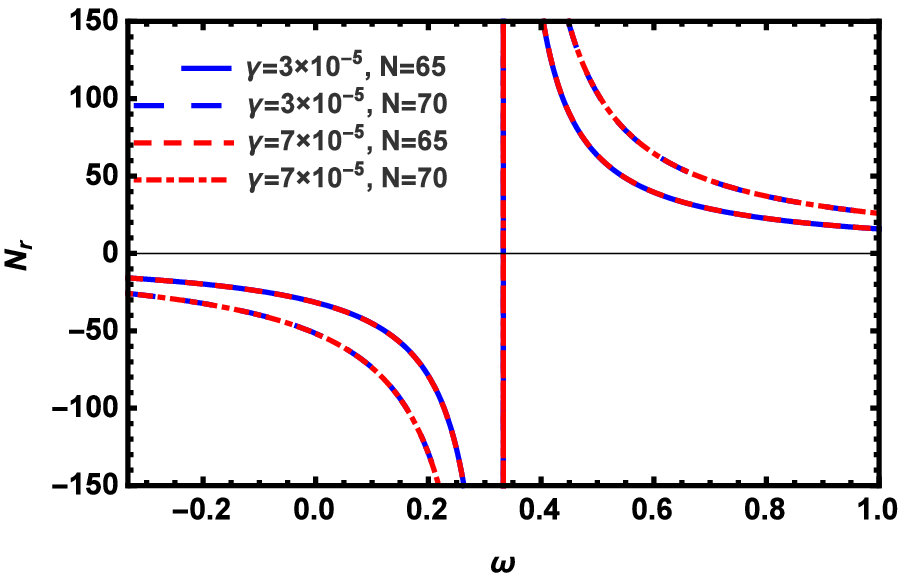}}
        \subfigure[]{\includegraphics[width=7cm]{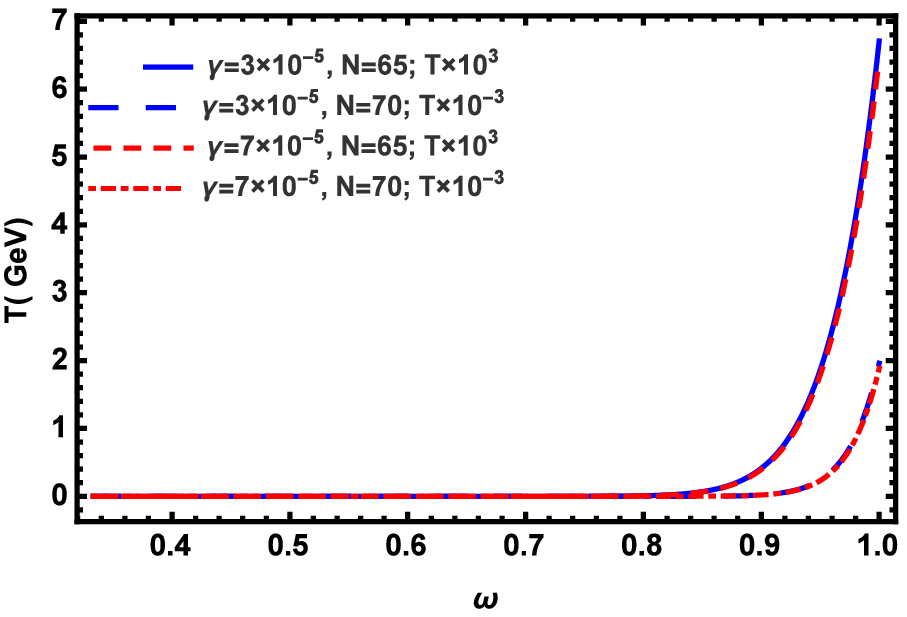}}
        \caption{The behavior of $N_r$ and $T_r$ versus the equation of the state parameter $\omega$ for different values of $\gamma$ and $N$ taken from parametric space Fig.\ref{TmodegN}. }\label{TNreheatingTmode}
      \end{figure}
      %%%%%%%%%%%%%%%%%%%%%%%%%%%%%%%%%%%%%%%%%%%%%%%%%%%%%%%%%%%
      The temperature $T_r$ is more sensitive to the value of the $N$ that the constant $\gamma$ in which by increasing $N$, the temperature decreases as well. For instance, for $N=65$ the temperature $T_r$ is about $10^{3} \; {\rm GeV}$, but for $N=70$ the temperature decrease to the order $10^{-3} \; {\rm GeV}$. \\
      The reheating temperature in the T-mode potential is again more sensitive to the number of e-fold, however, for this case it decrease by reduction of N. The temperature $T_r$ crosses the lower bond for higher values of $N$ in which for $N=70$ it stands on the lower edge of the range of $T_r$. In contrast to the previous case, the reheating temperature could impose a higher limit of the number of e-fold in which above this limit the predicted $T_r$ will be out of the mentioned range.
  \item \textbf{Generalized T-mode potential case:} As the last case in our study, $N_r$ and $T_r$ are illustrated versus the parameter $\omega$ in Fig.\ref{TNreheatingGTmode} for different values of $\gamma$ and $N$ taken from Fig.\ref{gNGT}. Again, $N_r$ is negative for $-1/3 < \omega < 1/3$ which is not acceptable. It is positive for $1/3 < \omega < 1$ and gets smaller as $\omega$ approaches one. \\
      %%%%%%%%%%%%%%%%%%%%%%%%%%%%%%%%%%%%%%%%%%%%%%%%%%%%%%%%%%
      \begin{figure}[h]
        \centering
        \subfigure[]{\includegraphics[width=7.5cm]{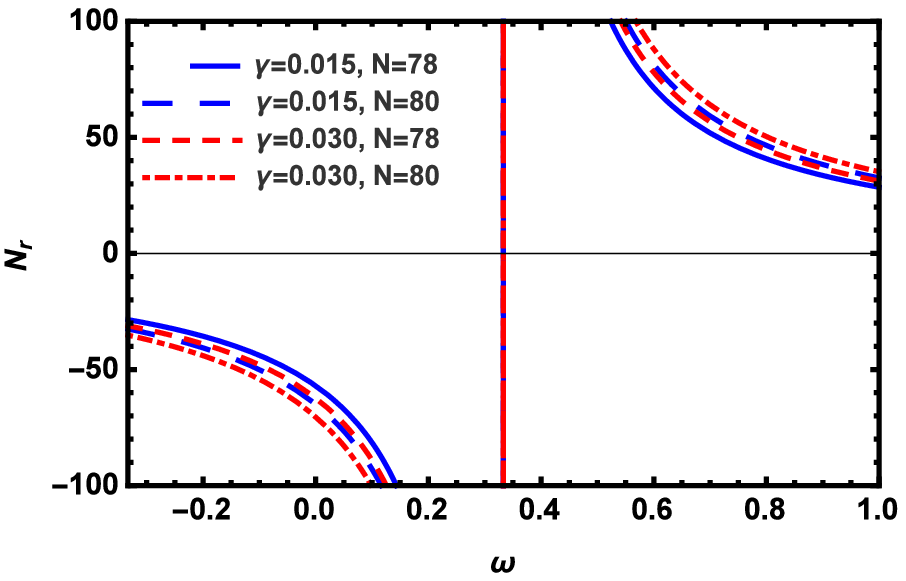}}
        \subfigure[]{\includegraphics[width=7cm]{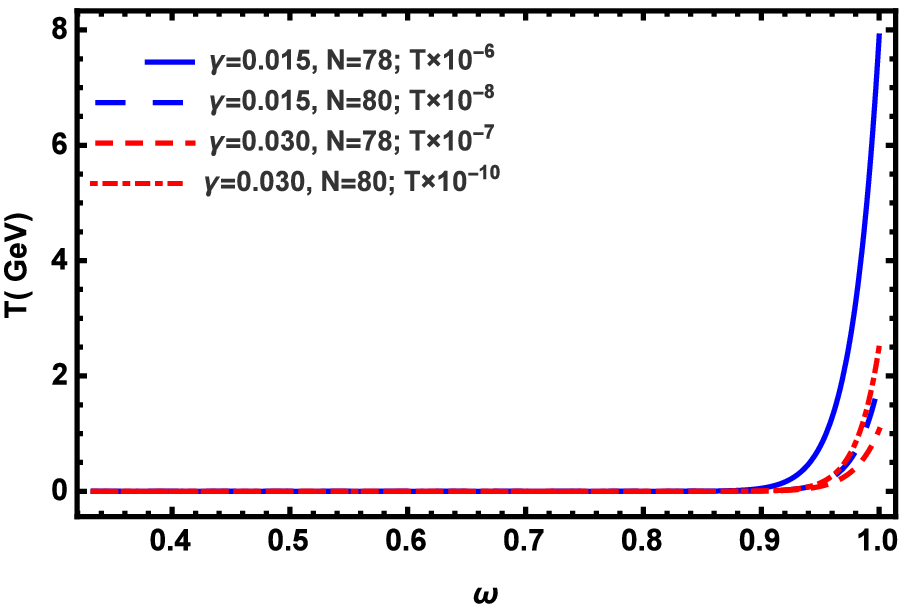}}
        \caption{The behavior of $N_r$ and $T_r$ versus the equation of the state parameter $\omega$ for different values of $\gamma$ and $N$ taken from parametric space Fig.\ref{gNGT}. }\label{TNreheatingGTmode}
      \end{figure}
      %%%%%%%%%%%%%%%%%%%%%%%%%%%%%%%%%%%%%%%%%%%%%%%%%%%%%%%%%%%
      The estimated reheating temperature for the chosen values of $\gamma$ and $N$ indicates that $T_r$ decreases by enhancement of $N$ and $\gamma$. Also, it seems that $T_r$ for the case is very small of the order of $10^{-8}$ and even smaller. \\
      For the last case, the reheating temperature seems to be completely out of the range. None of the obtained values of the constants in Fig.\ref{gNGT} could predict a $T_r$ in the range.
\end{itemize}

%=============================================================%
%=============================================================%
%=============================================================%
%=============================================================%
%=======================  Section 5 ==========================%
%=============================================================%
%=============================================================%
%=============================================================%
%=============================================================%
\section{Trans-Planckian Censorship Conjecture}\label{TransPlanckian}
The recently proposed Trans-Planckian censorship conjecture (TCC) states that the consistent theory of quantum gravity does not allow any fluctuation with wavelength equal to or shorter than the Planck length be stretched and exit the horizon and turn to a classical fluctuation. This conjecture leads to the following condition
\begin{equation}\label{TCC}
  e^N = {a_e \over a_i} < {M_p \over H_e}.
\end{equation}
%It is expected that the inflation solve the problem of structure formation which requires that the current comoving Hubble radius stands inside the comoving Hubble radius at the onset of inflation, i.e. $\left( a_0 H_0 \right)^{-1}  \leq \left( a_i H_i \right)^{-1}$. This condition is read as
%\begin{equation}\label{structurecondition}
%  H_0^{-1} \; e^{-N} \; {a_e \over a_r} \; {a_r \over a_0} \leq H_i^{-1}
%\end{equation}
where $H_e$ is the Hubble parameter at the end of inflation, $a_i$ is the scale factor at the beginning of inflation, and $a_e$ is the scale factor at the end of inflation. The main goal of this section is to consider whether the above condition, known as TCC, could be satisfied by the presented model. \\ 
The first step is to evalute the Hubble parameter $H_e$. From the Friedmann equation \eqref{Friedmann}, it is reazlized that to evaluate the Hubble parameter is proportional to the energy density $\rho$. Then, the above condition could be rewritten as\footnote{Remember that it is was assumed that inflation occurs in high energy regime, where $\rho \gg \lambda$. Then, the quadratic term of energy density dominates over the linear term.}  
\begin{equation}\label{TCCFriedmann}
  e^N < {3 m_p M_5^3 \over 4\pi \rho_e}.
\end{equation}
The energy density is a comibination of the kinetic term $\dot\phi^2$ and the potential $V(\phi)$. To calculate $\rho_e$, both kinetic term and the potential depend are required to obtained at the end of inflation. \\
Let consider the first case of the model. Assume that the inflation ends after $N=60$ e-fold. Then, from Fig.(\ref{nN}), it is realized that the free constant $n$ could not take any value. In fact, in Sec.\ref{ObservationSwampland}, we found a range for the paramter $n$ which brings the model to a good consistency with observational data and simultaneously satisfies the swampland criteria. Now, we want to know if this range of $n$ could also satisfy the TCC. The next parameter which is effective in the TCC is the five-dimensional Planck mass $M_5$. For the first case, this parameter was taken as $M_5 = 2 \times 10^{14}$. Then, if one takes $N=60$ and $M_5 = 2 \times 10^{14}$, the condition \eqref{TCCFriedmann} leads to  the following condition
\begin{equation}
\rho_e < 5.04 \times 10^{35}  {\rm GeV^4}
\end{equation}
It means that the energy density at the end of inflation should be smaller than the order of $10^{35}  {\rm GeV^4}$. However , the result of the first case for the energy density shows that $\rho_e$ is much larger than this value. Another impostant point is that this value for the energy density is in direct tension with our assumption that the inflation occurs in high energy regime.  \\
Therefore, it sounds that the TCC will not be satisfied by the model. But, there is a point which is worth mentioning. The value of the five dimensional Planck mass is very effecctive here. If one takes higher values for $M_5$, the situation gets worse, and the brane tension $\lambda$ gets much larger than the energy density and the high energy regime assumption will be broken again. On the other hand, if one smaller values for $M_5$ there is a chance to satisfy the TCC. For instance, for $M_5=10^8 {\rm GeV}$, the TCC impose a condition on the energy density as $\rho_e < 10^{17} {\rm GeV^4}$, where the brane tension is about $10^{12}  {\rm GeV^4}$. Then, by satisfying the TCC, the high energy regime assumption will not be broken. But the point is that for this value of $M_5$, the energy scale of inflation decreases, because smaller values $M_5$ leads to smaller potential at the beginning of inflation. Therefore, although the smaller $M_5$ satisfy the TCC, it results in small energy scale for inflation.

%=============================================================%
%=============================================================%
%=============================================================%
%=============================================================%
%=======================  Section 5 ==========================%
%=============================================================%
%=============================================================%
%=============================================================%
%=============================================================%
\section{Conclusion}
We studied the inflationary  scenario in the frame work  of brane gravity, where all   standard particle live on a four-dimensional space-time embedded in five-dimensional space-time. In particular, the inflaton is confined on the brane  and  its energy density dominates the  universe. Unlike  in the standard cosmology,  the Friedmann equation contains a term quadratic  in the energy density which affects the dynamics  in  the high energy  regime. After deriving the general expressions of the slow-roll parameters and the density perturbations generated during inflation, we investigated in details  some  well  known class of inflaton potentials. Instead of comparing the result for some random values of potential parameters (which are constants), a programming code was utilized to find the best values for the parameter. In this regard, by comparing the predicted $r$ and $n_s$ of the model with the $r-n_s$ diagram of Planck we could illustrate an allowed range of the parameters in that yield values of the spectral index and the tensor-to-scalar ratio that are in agreement with the Planck data for every point in the range. We also showed that these type of potentials satisfy the swampland criteria. \\
Reheating is a necessary phase which inseparable for any (cold) inflationary scenario which warms up the universe. The final temperature, reheating temperature, is required to stand in the range $1 \; {\rm MeV} < T_r < 10^{9-10} \; {\rm GeV}$ in order to recover the successful hot big bang nucleosynthesis and on the other hand to avoid the reproduction of any unwanted particle. Then, considering the reheating phase for the model could be count as a good way of constraining the parameter. Our consideration indicated that the estimated reheating temperature goes above the range for power-law and natural potentials. Therefore, the allowed range of reheating temperature could be applied to put another constraint for the model which clearly limit the allowed values of the potential parameter. For the T-mode potential there is almost a reverse situation in which the predicted reheating temperature could goes below the range. However, the range of reheating temperature could be used to put more constraint of the potential parameter. The result for the generalized T-mode potential is clear, it is absolutely out of allowed range of reheating temperature. Then it could not be counted as a good model of inflation.  \\
The TCC was considered in the last section as another way of confining the model. The conjecture states that the mode with wavelength shorter than the Planck scale never cross the Hubble horizon. It was explained that to satisfy the condition the five-dimensional Planck mass (or the brane tension) should accept an upper bound. On the other hand, lower values of $M_5$ leads to lower energy scale for inflation. The conjecture in the brane inflation leads to strong bound on the potential which might not be preserved.

%=============================================================%
%=============================================================%
%=======================  References =========================%
%=============================================================%
%=============================================================%



%% References


%\bibliography{sample.bib}



%\begin{thebibliography}{00}

\bibliography{35Refbib}

\begin{thebibliography}{111}
\expandafter\ifx\csname natexlab\endcsname\relax\def\natexlab#1{#1}\fi
\expandafter\ifx\csname bibnamefont\endcsname\relax
  \def\bibnamefont#1{#1}\fi
\expandafter\ifx\csname bibfnamefont\endcsname\relax
  \def\bibfnamefont#1{#1}\fi
\expandafter\ifx\csname citenamefont\endcsname\relax
  \def\citenamefont#1{#1}\fi
\expandafter\ifx\csname url\endcsname\relax
  \def\url#1{\texttt{#1}}\fi
\expandafter\ifx\csname urlprefix\endcsname\relax\def\urlprefix{URL }\fi
\providecommand{\bibinfo}[2]{#2}
\providecommand{\eprint}[2][]{\url{#2}}

\bibitem[{\citenamefont{Ade et~al.}(2014)}]{Planck:2013jfk}
\bibinfo{author}{\bibfnamefont{P.~A.~R.} \bibnamefont{Ade}}
  \bibnamefont{et~al.} (\bibinfo{collaboration}{Planck}),
  \bibinfo{journal}{Astron. Astrophys.} \textbf{\bibinfo{volume}{571}},
  \bibinfo{pages}{A22} (\bibinfo{year}{2014}), \eprint{arXiv:1303.5082}.

\bibitem[{\citenamefont{Ade et~al.}(2016)}]{Ade:2015lrj}
\bibinfo{author}{\bibfnamefont{P.~A.~R.} \bibnamefont{Ade}}
  \bibnamefont{et~al.} (\bibinfo{collaboration}{Planck}),
  \bibinfo{journal}{Astron. Astrophys.} \textbf{\bibinfo{volume}{594}},
  \bibinfo{pages}{A20} (\bibinfo{year}{2016}), \eprint{arXiv:1502.02114}.

\bibitem[{\citenamefont{Akrami et~al.}(2018)}]{Akrami:2018odb}
\bibinfo{author}{\bibfnamefont{Y.}~\bibnamefont{Akrami}} \bibnamefont{et~al.}
  (\bibinfo{collaboration}{Planck}) (\bibinfo{year}{2018}),
  \eprint{arXiv:1807.06211}.

\bibitem[{\citenamefont{Starobinsky}(1980)}]{starobinsky1980new}
\bibinfo{author}{\bibfnamefont{A.~A.} \bibnamefont{Starobinsky}},
  \bibinfo{journal}{Physics Letters B} \textbf{\bibinfo{volume}{91}},
  \bibinfo{pages}{99} (\bibinfo{year}{1980}).

\bibitem[{\citenamefont{Guth}(1981)}]{Guth:1980zm}
\bibinfo{author}{\bibfnamefont{A.~H.} \bibnamefont{Guth}},
  \bibinfo{journal}{Phys. Rev.} \textbf{\bibinfo{volume}{D23}},
  \bibinfo{pages}{347} (\bibinfo{year}{1981}), \bibinfo{note}{[Adv. Ser.
  Astrophys. Cosmol.3,139(1987)]}.

\bibitem[{\citenamefont{Albrecht and Steinhardt}(1982)}]{albrecht1982cosmology}
\bibinfo{author}{\bibfnamefont{A.}~\bibnamefont{Albrecht}} \bibnamefont{and}
  \bibinfo{author}{\bibfnamefont{P.~J.} \bibnamefont{Steinhardt}},
  \bibinfo{journal}{Physical Review Letters} \textbf{\bibinfo{volume}{48}},
  \bibinfo{pages}{1220} (\bibinfo{year}{1982}).

\bibitem[{\citenamefont{Linde}(1982)}]{linde1982new}
\bibinfo{author}{\bibfnamefont{A.~D.} \bibnamefont{Linde}},
  \bibinfo{journal}{Physics Letters B} \textbf{\bibinfo{volume}{108}},
  \bibinfo{pages}{389} (\bibinfo{year}{1982}).

\bibitem[{\citenamefont{Linde}(1983)}]{linde1983chaotic}
\bibinfo{author}{\bibfnamefont{A.~D.} \bibnamefont{Linde}},
  \bibinfo{journal}{Physics Letters B} \textbf{\bibinfo{volume}{129}},
  \bibinfo{pages}{177} (\bibinfo{year}{1983}).

\bibitem[{\citenamefont{Barenboim and Kinney}(2007)}]{Barenboim:2007ii}
\bibinfo{author}{\bibfnamefont{G.}~\bibnamefont{Barenboim}} \bibnamefont{and}
  \bibinfo{author}{\bibfnamefont{W.~H.} \bibnamefont{Kinney}},
  \bibinfo{journal}{JCAP} \textbf{\bibinfo{volume}{0703}}, \bibinfo{pages}{014}
  (\bibinfo{year}{2007}), \eprint{astro-ph/0701343}.

\bibitem[{\citenamefont{Franche et~al.}(2010)\citenamefont{Franche, Gwyn,
  Underwood, and Wissanji}}]{Franche:2010yj}
\bibinfo{author}{\bibfnamefont{P.}~\bibnamefont{Franche}},
  \bibinfo{author}{\bibfnamefont{R.}~\bibnamefont{Gwyn}},
  \bibinfo{author}{\bibfnamefont{B.}~\bibnamefont{Underwood}},
  \bibnamefont{and} \bibinfo{author}{\bibfnamefont{A.}~\bibnamefont{Wissanji}},
  \bibinfo{journal}{Phys. Rev.} \textbf{\bibinfo{volume}{D82}},
  \bibinfo{pages}{063528} (\bibinfo{year}{2010}), \eprint{arXiv:1002.2639}.

\bibitem[{\citenamefont{Unnikrishnan et~al.}(2012)\citenamefont{Unnikrishnan,
  Sahni, and Toporensky}}]{Unnikrishnan:2012zu}
\bibinfo{author}{\bibfnamefont{S.}~\bibnamefont{Unnikrishnan}},
  \bibinfo{author}{\bibfnamefont{V.}~\bibnamefont{Sahni}}, \bibnamefont{and}
  \bibinfo{author}{\bibfnamefont{A.}~\bibnamefont{Toporensky}},
  \bibinfo{journal}{JCAP} \textbf{\bibinfo{volume}{1208}}, \bibinfo{pages}{018}
  (\bibinfo{year}{2012}), \eprint{arXiv:1205.0786}.

\bibitem[{\citenamefont{Gwyn et~al.}(2013)\citenamefont{Gwyn, Rummel, and
  Westphal}}]{Gwyn:2012ey}
\bibinfo{author}{\bibfnamefont{R.}~\bibnamefont{Gwyn}},
  \bibinfo{author}{\bibfnamefont{M.}~\bibnamefont{Rummel}}, \bibnamefont{and}
  \bibinfo{author}{\bibfnamefont{A.}~\bibnamefont{Westphal}},
  \bibinfo{journal}{JCAP} \textbf{\bibinfo{volume}{1312}}, \bibinfo{pages}{010}
  (\bibinfo{year}{2013}), \eprint{arXiv:1212.4135}.

\bibitem[{\citenamefont{Rezazadeh et~al.}(2015)\citenamefont{Rezazadeh, Karami,
  and Karimi}}]{Rezazadeh:2014fwa}
\bibinfo{author}{\bibfnamefont{K.}~\bibnamefont{Rezazadeh}},
  \bibinfo{author}{\bibfnamefont{K.}~\bibnamefont{Karami}}, \bibnamefont{and}
  \bibinfo{author}{\bibfnamefont{P.}~\bibnamefont{Karimi}},
  \bibinfo{journal}{JCAP} \textbf{\bibinfo{volume}{1509}}, \bibinfo{pages}{053}
  (\bibinfo{year}{2015}), \eprint{arXiv:1411.7302}.

\bibitem[{\citenamefont{Céspedes and Davis}(2015)}]{Cespedes:2015jga}
\bibinfo{author}{\bibfnamefont{S.}~\bibnamefont{Céspedes}} \bibnamefont{and}
  \bibinfo{author}{\bibfnamefont{A.-C.} \bibnamefont{Davis}},
  \bibinfo{journal}{JCAP} \textbf{\bibinfo{volume}{1511}}, \bibinfo{pages}{014}
  (\bibinfo{year}{2015}), \eprint{arXiv:1506.01244}.

\bibitem[{\citenamefont{Stein and Kinney}(2017)}]{Stein:2016jja}
\bibinfo{author}{\bibfnamefont{N.~K.} \bibnamefont{Stein}} \bibnamefont{and}
  \bibinfo{author}{\bibfnamefont{W.~H.} \bibnamefont{Kinney}},
  \bibinfo{journal}{JCAP} \textbf{\bibinfo{volume}{1704}}, \bibinfo{pages}{006}
  (\bibinfo{year}{2017}), \eprint{arXiv:1609.08959}.

\bibitem[{\citenamefont{Pinhero and Pal}(2017)}]{Pinhero:2017lni}
\bibinfo{author}{\bibfnamefont{T.}~\bibnamefont{Pinhero}} \bibnamefont{and}
  \bibinfo{author}{\bibfnamefont{S.}~\bibnamefont{Pal}} (\bibinfo{year}{2017}),
  \eprint{arXiv:1703.07165}.

\bibitem[{\citenamefont{Fairbairn and Tytgat}(2002)}]{Fairbairn:2002yp}
\bibinfo{author}{\bibfnamefont{M.}~\bibnamefont{Fairbairn}} \bibnamefont{and}
  \bibinfo{author}{\bibfnamefont{M.~H.~G.} \bibnamefont{Tytgat}},
  \bibinfo{journal}{Phys. Lett.} \textbf{\bibinfo{volume}{B546}},
  \bibinfo{pages}{1} (\bibinfo{year}{2002}), \eprint{hep-th/0204070}.

\bibitem[{\citenamefont{Mukohyama}(2002)}]{Mukohyama:2002cn}
\bibinfo{author}{\bibfnamefont{S.}~\bibnamefont{Mukohyama}},
  \bibinfo{journal}{Phys. Rev.} \textbf{\bibinfo{volume}{D66}},
  \bibinfo{pages}{024009} (\bibinfo{year}{2002}), \eprint{hep-th/0204084}.

\bibitem[{\citenamefont{Feinstein}(2002)}]{Feinstein:2002aj}
\bibinfo{author}{\bibfnamefont{A.}~\bibnamefont{Feinstein}},
  \bibinfo{journal}{Phys. Rev.} \textbf{\bibinfo{volume}{D66}},
  \bibinfo{pages}{063511} (\bibinfo{year}{2002}), \eprint{hep-th/0204140}.

\bibitem[{\citenamefont{Padmanabhan}(2002)}]{Padmanabhan:2002cp}
\bibinfo{author}{\bibfnamefont{T.}~\bibnamefont{Padmanabhan}},
  \bibinfo{journal}{Phys. Rev.} \textbf{\bibinfo{volume}{D66}},
  \bibinfo{pages}{021301} (\bibinfo{year}{2002}), \eprint{hep-th/0204150}.

\bibitem[{\citenamefont{Spalinski}(2007)}]{Spalinski:2007dv}
\bibinfo{author}{\bibfnamefont{M.}~\bibnamefont{Spalinski}},
  \bibinfo{journal}{JCAP} \textbf{\bibinfo{volume}{0705}}, \bibinfo{pages}{017}
  (\bibinfo{year}{2007}), \eprint{hep-th/0702196}.

\bibitem[{\citenamefont{Bessada et~al.}(2009)\citenamefont{Bessada, Kinney, and
  Tzirakis}}]{Bessada:2009pe}
\bibinfo{author}{\bibfnamefont{D.}~\bibnamefont{Bessada}},
  \bibinfo{author}{\bibfnamefont{W.~H.} \bibnamefont{Kinney}},
  \bibnamefont{and} \bibinfo{author}{\bibfnamefont{K.}~\bibnamefont{Tzirakis}},
  \bibinfo{journal}{JCAP} \textbf{\bibinfo{volume}{0909}}, \bibinfo{pages}{031}
  (\bibinfo{year}{2009}), \eprint{arXiv:0907.1311}.

\bibitem[{\citenamefont{Weller et~al.}(2012)\citenamefont{Weller, van~de Bruck,
  and Mota}}]{Weller:2011ey}
\bibinfo{author}{\bibfnamefont{J.~M.} \bibnamefont{Weller}},
  \bibinfo{author}{\bibfnamefont{C.}~\bibnamefont{van~de Bruck}},
  \bibnamefont{and} \bibinfo{author}{\bibfnamefont{D.~F.} \bibnamefont{Mota}},
  \bibinfo{journal}{JCAP} \textbf{\bibinfo{volume}{1206}}, \bibinfo{pages}{002}
  (\bibinfo{year}{2012}), \eprint{arXiv:1111.0237}.

\bibitem[{\citenamefont{Nazavari et~al.}(2016)\citenamefont{Nazavari,
  Mohammadi, Ossoulian, and Saaidi}}]{Nazavari:2016yaa}
\bibinfo{author}{\bibfnamefont{N.}~\bibnamefont{Nazavari}},
  \bibinfo{author}{\bibfnamefont{A.}~\bibnamefont{Mohammadi}},
  \bibinfo{author}{\bibfnamefont{Z.}~\bibnamefont{Ossoulian}},
  \bibnamefont{and} \bibinfo{author}{\bibfnamefont{K.}~\bibnamefont{Saaidi}},
  \bibinfo{journal}{Phys. Rev.} \textbf{\bibinfo{volume}{D93}},
  \bibinfo{pages}{123504} (\bibinfo{year}{2016}), \eprint{arXiv:1708.03676}.

\bibitem[{\citenamefont{Amani et~al.}(2018)\citenamefont{Amani, Rezazadeh,
  Abdolmaleki, and Karami}}]{Amani:2018ueu}
\bibinfo{author}{\bibfnamefont{R.}~\bibnamefont{Amani}},
  \bibinfo{author}{\bibfnamefont{K.}~\bibnamefont{Rezazadeh}},
  \bibinfo{author}{\bibfnamefont{A.}~\bibnamefont{Abdolmaleki}},
  \bibnamefont{and} \bibinfo{author}{\bibfnamefont{K.}~\bibnamefont{Karami}},
  \bibinfo{journal}{Astrophys. J.} \textbf{\bibinfo{volume}{853}},
  \bibinfo{pages}{188} (\bibinfo{year}{2018}), \eprint{arXiv:1802.06075}.

\bibitem[{\citenamefont{Golanbari et~al.}(2020)\citenamefont{Golanbari,
  Mohammadi, and Saaidi}}]{Mohammadi:2018zkf}
\bibinfo{author}{\bibfnamefont{T.}~\bibnamefont{Golanbari}},
  \bibinfo{author}{\bibfnamefont{A.}~\bibnamefont{Mohammadi}},
  \bibnamefont{and} \bibinfo{author}{\bibfnamefont{K.}~\bibnamefont{Saaidi}},
  \bibinfo{journal}{Phys. Dark Univ.} \textbf{\bibinfo{volume}{27}},
  \bibinfo{pages}{100456} (\bibinfo{year}{2020}), \eprint{arXiv:1808.07246}.

\bibitem[{\citenamefont{Maeda and Yamamoto}(2013)}]{maeda2013stability}
\bibinfo{author}{\bibfnamefont{K.-i.} \bibnamefont{Maeda}} \bibnamefont{and}
  \bibinfo{author}{\bibfnamefont{K.}~\bibnamefont{Yamamoto}},
  \bibinfo{journal}{Journal of Cosmology and Astroparticle Physics}
  \textbf{\bibinfo{volume}{2013}}, \bibinfo{pages}{018} (\bibinfo{year}{2013}).

\bibitem[{\citenamefont{Abolhasani et~al.}(2014)\citenamefont{Abolhasani,
  Emami, and Firouzjahi}}]{abolhasani2014primordial}
\bibinfo{author}{\bibfnamefont{A.~A.} \bibnamefont{Abolhasani}},
  \bibinfo{author}{\bibfnamefont{R.}~\bibnamefont{Emami}}, \bibnamefont{and}
  \bibinfo{author}{\bibfnamefont{H.}~\bibnamefont{Firouzjahi}},
  \bibinfo{journal}{Journal of Cosmology and Astroparticle Physics}
  \textbf{\bibinfo{volume}{2014}}, \bibinfo{pages}{016} (\bibinfo{year}{2014}).

\bibitem[{\citenamefont{Alexander et~al.}(2015)\citenamefont{Alexander, Jyoti,
  Kosowsky, and Marcian{\`o}}}]{alexander2015dynamics}
\bibinfo{author}{\bibfnamefont{S.}~\bibnamefont{Alexander}},
  \bibinfo{author}{\bibfnamefont{D.}~\bibnamefont{Jyoti}},
  \bibinfo{author}{\bibfnamefont{A.}~\bibnamefont{Kosowsky}}, \bibnamefont{and}
  \bibinfo{author}{\bibfnamefont{A.}~\bibnamefont{Marcian{\`o}}},
  \bibinfo{journal}{Journal of Cosmology and Astroparticle Physics}
  \textbf{\bibinfo{volume}{2015}}, \bibinfo{pages}{005} (\bibinfo{year}{2015}).

\bibitem[{\citenamefont{Tirandari and Saaidi}(2017)}]{tirandari2017anisotropic}
\bibinfo{author}{\bibfnamefont{M.}~\bibnamefont{Tirandari}} \bibnamefont{and}
  \bibinfo{author}{\bibfnamefont{K.}~\bibnamefont{Saaidi}},
  \bibinfo{journal}{Nuclear Physics B} \textbf{\bibinfo{volume}{925}},
  \bibinfo{pages}{403} (\bibinfo{year}{2017}).

\bibitem[{\citenamefont{Berera}(1995)}]{berera1995warm}
\bibinfo{author}{\bibfnamefont{A.}~\bibnamefont{Berera}},
  \bibinfo{journal}{Physical Review Letters} \textbf{\bibinfo{volume}{75}},
  \bibinfo{pages}{3218} (\bibinfo{year}{1995}).

\bibitem[{\citenamefont{Berera}(2000)}]{berera2000warm}
\bibinfo{author}{\bibfnamefont{A.}~\bibnamefont{Berera}},
  \bibinfo{journal}{Nuclear Physics B} \textbf{\bibinfo{volume}{585}},
  \bibinfo{pages}{666} (\bibinfo{year}{2000}).

\bibitem[{\citenamefont{Taylor and Berera}(2000)}]{taylor2000perturbation}
\bibinfo{author}{\bibfnamefont{A.}~\bibnamefont{Taylor}} \bibnamefont{and}
  \bibinfo{author}{\bibfnamefont{A.}~\bibnamefont{Berera}},
  \bibinfo{journal}{Physical Review D} \textbf{\bibinfo{volume}{62}},
  \bibinfo{pages}{083517} (\bibinfo{year}{2000}).

\bibitem[{\citenamefont{Hall et~al.}(2004)\citenamefont{Hall, Moss, and
  Berera}}]{hall2004scalar}
\bibinfo{author}{\bibfnamefont{L.~M.} \bibnamefont{Hall}},
  \bibinfo{author}{\bibfnamefont{I.~G.} \bibnamefont{Moss}}, \bibnamefont{and}
  \bibinfo{author}{\bibfnamefont{A.}~\bibnamefont{Berera}},
  \bibinfo{journal}{Physical Review D} \textbf{\bibinfo{volume}{69}},
  \bibinfo{pages}{083525} (\bibinfo{year}{2004}).

\bibitem[{\citenamefont{Bastero-Gil and Berera}(2005)}]{BasteroGil:2004tg}
\bibinfo{author}{\bibfnamefont{M.}~\bibnamefont{Bastero-Gil}} \bibnamefont{and}
  \bibinfo{author}{\bibfnamefont{A.}~\bibnamefont{Berera}},
  \bibinfo{journal}{Phys. Rev.} \textbf{\bibinfo{volume}{D71}},
  \bibinfo{pages}{063515} (\bibinfo{year}{2005}), \eprint{hep-ph/0411144}.

\bibitem[{\citenamefont{Bastero-Gil et~al.}(2016)\citenamefont{Bastero-Gil,
  Berera, Ramos, and Rosa}}]{Bastero-Gil:2016qru}
\bibinfo{author}{\bibfnamefont{M.}~\bibnamefont{Bastero-Gil}},
  \bibinfo{author}{\bibfnamefont{A.}~\bibnamefont{Berera}},
  \bibinfo{author}{\bibfnamefont{R.~O.} \bibnamefont{Ramos}}, \bibnamefont{and}
  \bibinfo{author}{\bibfnamefont{J.~G.} \bibnamefont{Rosa}},
  \bibinfo{journal}{Phys. Rev. Lett.} \textbf{\bibinfo{volume}{117}},
  \bibinfo{pages}{151301} (\bibinfo{year}{2016}), \eprint{1604.08838}.

\bibitem[{\citenamefont{Rosa and Ventura}(2019)}]{Rosa:2018iff}
\bibinfo{author}{\bibfnamefont{J.~G.} \bibnamefont{Rosa}} \bibnamefont{and}
  \bibinfo{author}{\bibfnamefont{L.~B.} \bibnamefont{Ventura}},
  \bibinfo{journal}{Phys. Rev. Lett.} \textbf{\bibinfo{volume}{122}},
  \bibinfo{pages}{161301} (\bibinfo{year}{2019}), \eprint{1811.05493}.

\bibitem[{\citenamefont{Bastero-Gil et~al.}(2019)\citenamefont{Bastero-Gil,
  Berera, Ramos, and Rosa}}]{Bastero-Gil:2019gao}
\bibinfo{author}{\bibfnamefont{M.}~\bibnamefont{Bastero-Gil}},
  \bibinfo{author}{\bibfnamefont{A.}~\bibnamefont{Berera}},
  \bibinfo{author}{\bibfnamefont{R.~O.} \bibnamefont{Ramos}}, \bibnamefont{and}
  \bibinfo{author}{\bibfnamefont{J.~G.} \bibnamefont{Rosa}}
  (\bibinfo{year}{2019}), \eprint{1907.13410}.

\bibitem[{\citenamefont{Sayar et~al.}(2017)\citenamefont{Sayar, Mohammadi,
  Akhtari, and Saaidi}}]{Sayar:2017pam}
\bibinfo{author}{\bibfnamefont{K.}~\bibnamefont{Sayar}},
  \bibinfo{author}{\bibfnamefont{A.}~\bibnamefont{Mohammadi}},
  \bibinfo{author}{\bibfnamefont{L.}~\bibnamefont{Akhtari}}, \bibnamefont{and}
  \bibinfo{author}{\bibfnamefont{K.}~\bibnamefont{Saaidi}},
  \bibinfo{journal}{Phys. Rev.} \textbf{\bibinfo{volume}{D95}},
  \bibinfo{pages}{023501} (\bibinfo{year}{2017}), \eprint{arXiv:1708.01714}.

\bibitem[{\citenamefont{Akhtari et~al.}(2017)\citenamefont{Akhtari, Mohammadi,
  Sayar, and Saaidi}}]{Akhtari:2017mxc}
\bibinfo{author}{\bibfnamefont{L.}~\bibnamefont{Akhtari}},
  \bibinfo{author}{\bibfnamefont{A.}~\bibnamefont{Mohammadi}},
  \bibinfo{author}{\bibfnamefont{K.}~\bibnamefont{Sayar}}, \bibnamefont{and}
  \bibinfo{author}{\bibfnamefont{K.}~\bibnamefont{Saaidi}},
  \bibinfo{journal}{Astropart. Phys.} \textbf{\bibinfo{volume}{90}},
  \bibinfo{pages}{28} (\bibinfo{year}{2017}), \eprint{arXiv:1710.05793}.

\bibitem[{\citenamefont{Sheikhahmadi et~al.}(2019)\citenamefont{Sheikhahmadi,
  Mohammadi, Aghamohammadi, Harko, Herrera, Corda, Abebe, and
  Saaidi}}]{Sheikhahmadi:2019gzs}
\bibinfo{author}{\bibfnamefont{H.}~\bibnamefont{Sheikhahmadi}},
  \bibinfo{author}{\bibfnamefont{A.}~\bibnamefont{Mohammadi}},
  \bibinfo{author}{\bibfnamefont{A.}~\bibnamefont{Aghamohammadi}},
  \bibinfo{author}{\bibfnamefont{T.}~\bibnamefont{Harko}},
  \bibinfo{author}{\bibfnamefont{R.}~\bibnamefont{Herrera}},
  \bibinfo{author}{\bibfnamefont{C.}~\bibnamefont{Corda}},
  \bibinfo{author}{\bibfnamefont{A.}~\bibnamefont{Abebe}}, \bibnamefont{and}
  \bibinfo{author}{\bibfnamefont{K.}~\bibnamefont{Saaidi}},
  \bibinfo{journal}{Eur. Phys. J.} \textbf{\bibinfo{volume}{C79}},
  \bibinfo{pages}{1038} (\bibinfo{year}{2019}), \eprint{arXiv:1907.10966}.

\bibitem[{\citenamefont{Riotto}(2003)}]{Riotto:2002yw}
\bibinfo{author}{\bibfnamefont{A.}~\bibnamefont{Riotto}},
  \bibinfo{journal}{ICTP Lect. Notes Ser.} \textbf{\bibinfo{volume}{14}},
  \bibinfo{pages}{317} (\bibinfo{year}{2003}), \eprint{hep-ph/0210162}.

\bibitem[{\citenamefont{Baumann}(2011)}]{Baumann:2009ds}
\bibinfo{author}{\bibfnamefont{D.}~\bibnamefont{Baumann}}, in
  \emph{\bibinfo{booktitle}{{Physics of the large and the small, TASI 09,
  proceedings of the Theoretical Advanced Study Institute in Elementary
  Particle Physics, Boulder, Colorado, USA, 1-26 June 2009}}}
  (\bibinfo{year}{2011}), pp. \bibinfo{pages}{523--686},
  \eprint{arXiv:0907.5424}.

\bibitem[{\citenamefont{Weinberg}(2008)}]{Weinberg:2008zzc}
\bibinfo{author}{\bibfnamefont{S.}~\bibnamefont{Weinberg}},
  \emph{\bibinfo{title}{{Cosmology}}} (\bibinfo{year}{2008}), ISBN
  \bibinfo{isbn}{9780198526827},
  \urlprefix\url{http://www.oup.com/uk/catalogue/?ci=9780198526827}.

\bibitem[{\citenamefont{Lyth and Liddle}(2009)}]{Lyth:2009zz}
\bibinfo{author}{\bibfnamefont{D.~H.} \bibnamefont{Lyth}} \bibnamefont{and}
  \bibinfo{author}{\bibfnamefont{A.~R.} \bibnamefont{Liddle}},
  \emph{\bibinfo{title}{{The primordial density perturbation: Cosmology,
  inflation and the origin of structure}}} (\bibinfo{year}{2009}),
  \urlprefix\url{http://www.cambridge.org/uk/catalogue/catalogue.asp?isbn=9780521828499}.

\bibitem[{\citenamefont{Abbott et~al.}(1982)\citenamefont{Abbott, Farhi, and
  Wise}}]{Abbott:1982hn}
\bibinfo{author}{\bibfnamefont{L.~F.} \bibnamefont{Abbott}},
  \bibinfo{author}{\bibfnamefont{E.}~\bibnamefont{Farhi}}, \bibnamefont{and}
  \bibinfo{author}{\bibfnamefont{M.~B.} \bibnamefont{Wise}},
  \bibinfo{journal}{Phys. Lett. B} \textbf{\bibinfo{volume}{117}},
  \bibinfo{pages}{29} (\bibinfo{year}{1982}).

\bibitem[{\citenamefont{Albrecht et~al.}(1982)\citenamefont{Albrecht,
  Steinhardt, Turner, and Wilczek}}]{Albrecht:1982mp}
\bibinfo{author}{\bibfnamefont{A.}~\bibnamefont{Albrecht}},
  \bibinfo{author}{\bibfnamefont{P.~J.} \bibnamefont{Steinhardt}},
  \bibinfo{author}{\bibfnamefont{M.~S.} \bibnamefont{Turner}},
  \bibnamefont{and} \bibinfo{author}{\bibfnamefont{F.}~\bibnamefont{Wilczek}},
  \bibinfo{journal}{Phys. Rev. Lett.} \textbf{\bibinfo{volume}{48}},
  \bibinfo{pages}{1437} (\bibinfo{year}{1982}).

\bibitem[{\citenamefont{Dolgov and Linde}(1982)}]{Dolgov:1982th}
\bibinfo{author}{\bibfnamefont{A.~D.} \bibnamefont{Dolgov}} \bibnamefont{and}
  \bibinfo{author}{\bibfnamefont{A.~D.} \bibnamefont{Linde}},
  \bibinfo{journal}{Phys. Lett. B} \textbf{\bibinfo{volume}{116}},
  \bibinfo{pages}{329} (\bibinfo{year}{1982}).

\bibitem[{\citenamefont{Dolgov and Kirilova}(1990)}]{Dolgov:1989us}
\bibinfo{author}{\bibfnamefont{A.~D.} \bibnamefont{Dolgov}} \bibnamefont{and}
  \bibinfo{author}{\bibfnamefont{D.~P.} \bibnamefont{Kirilova}},
  \bibinfo{journal}{Sov. J. Nucl. Phys.} \textbf{\bibinfo{volume}{51}},
  \bibinfo{pages}{172} (\bibinfo{year}{1990}).

\bibitem[{\citenamefont{Traschen and Brandenberger}(1990)}]{Traschen:1990sw}
\bibinfo{author}{\bibfnamefont{J.~H.} \bibnamefont{Traschen}} \bibnamefont{and}
  \bibinfo{author}{\bibfnamefont{R.~H.} \bibnamefont{Brandenberger}},
  \bibinfo{journal}{Phys. Rev. D} \textbf{\bibinfo{volume}{42}},
  \bibinfo{pages}{2491} (\bibinfo{year}{1990}).

\bibitem[{\citenamefont{Shtanov et~al.}(1995)\citenamefont{Shtanov, Traschen,
  and Brandenberger}}]{Shtanov:1994ce}
\bibinfo{author}{\bibfnamefont{Y.}~\bibnamefont{Shtanov}},
  \bibinfo{author}{\bibfnamefont{J.~H.} \bibnamefont{Traschen}},
  \bibnamefont{and} \bibinfo{author}{\bibfnamefont{R.~H.}
  \bibnamefont{Brandenberger}}, \bibinfo{journal}{Phys. Rev. D}
  \textbf{\bibinfo{volume}{51}}, \bibinfo{pages}{5438} (\bibinfo{year}{1995}),
  \eprint{hep-ph/9407247}.

\bibitem[{\citenamefont{Kofman et~al.}(1994)\citenamefont{Kofman, Linde, and
  Starobinsky}}]{Kofman:1994rk}
\bibinfo{author}{\bibfnamefont{L.}~\bibnamefont{Kofman}},
  \bibinfo{author}{\bibfnamefont{A.~D.} \bibnamefont{Linde}}, \bibnamefont{and}
  \bibinfo{author}{\bibfnamefont{A.~A.} \bibnamefont{Starobinsky}},
  \bibinfo{journal}{Phys. Rev. Lett.} \textbf{\bibinfo{volume}{73}},
  \bibinfo{pages}{3195} (\bibinfo{year}{1994}), \eprint{hep-th/9405187}.

\bibitem[{\citenamefont{Kofman et~al.}(1997)\citenamefont{Kofman, Linde, and
  Starobinsky}}]{Kofman:1997yn}
\bibinfo{author}{\bibfnamefont{L.}~\bibnamefont{Kofman}},
  \bibinfo{author}{\bibfnamefont{A.~D.} \bibnamefont{Linde}}, \bibnamefont{and}
  \bibinfo{author}{\bibfnamefont{A.~A.} \bibnamefont{Starobinsky}},
  \bibinfo{journal}{Phys. Rev. D} \textbf{\bibinfo{volume}{56}},
  \bibinfo{pages}{3258} (\bibinfo{year}{1997}), \eprint{hep-ph/9704452}.

\bibitem[{\citenamefont{Bassett et~al.}(2006)\citenamefont{Bassett, Tsujikawa,
  and Wands}}]{Bassett:2005xm}
\bibinfo{author}{\bibfnamefont{B.~A.} \bibnamefont{Bassett}},
  \bibinfo{author}{\bibfnamefont{S.}~\bibnamefont{Tsujikawa}},
  \bibnamefont{and} \bibinfo{author}{\bibfnamefont{D.}~\bibnamefont{Wands}},
  \bibinfo{journal}{Rev. Mod. Phys.} \textbf{\bibinfo{volume}{78}},
  \bibinfo{pages}{537} (\bibinfo{year}{2006}), \eprint{astro-ph/0507632}.

\bibitem[{\citenamefont{Allahverdi et~al.}(2010)\citenamefont{Allahverdi,
  Brandenberger, Cyr-Racine, and Mazumdar}}]{Allahverdi:2010xz}
\bibinfo{author}{\bibfnamefont{R.}~\bibnamefont{Allahverdi}},
  \bibinfo{author}{\bibfnamefont{R.}~\bibnamefont{Brandenberger}},
  \bibinfo{author}{\bibfnamefont{F.-Y.} \bibnamefont{Cyr-Racine}},
  \bibnamefont{and} \bibinfo{author}{\bibfnamefont{A.}~\bibnamefont{Mazumdar}},
  \bibinfo{journal}{Ann. Rev. Nucl. Part. Sci.} \textbf{\bibinfo{volume}{60}},
  \bibinfo{pages}{27} (\bibinfo{year}{2010}), \eprint{1001.2600}.

\bibitem[{\citenamefont{Amin et~al.}(2014)\citenamefont{Amin, Hertzberg,
  Kaiser, and Karouby}}]{Amin:2014eta}
\bibinfo{author}{\bibfnamefont{M.~A.} \bibnamefont{Amin}},
  \bibinfo{author}{\bibfnamefont{M.~P.} \bibnamefont{Hertzberg}},
  \bibinfo{author}{\bibfnamefont{D.~I.} \bibnamefont{Kaiser}},
  \bibnamefont{and} \bibinfo{author}{\bibfnamefont{J.}~\bibnamefont{Karouby}},
  \bibinfo{journal}{Int. J. Mod. Phys. D} \textbf{\bibinfo{volume}{24}},
  \bibinfo{pages}{1530003} (\bibinfo{year}{2014}), \eprint{1410.3808}.

\bibitem[{\citenamefont{Randall and
  Sundrum}(1999{\natexlab{a}})}]{Randall:1999ee}
\bibinfo{author}{\bibfnamefont{L.}~\bibnamefont{Randall}} \bibnamefont{and}
  \bibinfo{author}{\bibfnamefont{R.}~\bibnamefont{Sundrum}},
  \bibinfo{journal}{Phys. Rev. Lett.} \textbf{\bibinfo{volume}{83}},
  \bibinfo{pages}{3370} (\bibinfo{year}{1999}{\natexlab{a}}),
  \eprint{hep-ph/9905221}.

\bibitem[{\citenamefont{Randall and
  Sundrum}(1999{\natexlab{b}})}]{Randall:1999vf}
\bibinfo{author}{\bibfnamefont{L.}~\bibnamefont{Randall}} \bibnamefont{and}
  \bibinfo{author}{\bibfnamefont{R.}~\bibnamefont{Sundrum}},
  \bibinfo{journal}{Phys. Rev. Lett.} \textbf{\bibinfo{volume}{83}},
  \bibinfo{pages}{4690} (\bibinfo{year}{1999}{\natexlab{b}}),
  \eprint{hep-th/9906064}.

\bibitem[{\citenamefont{Maartens et~al.}(2000)\citenamefont{Maartens, Wands,
  Bassett, and Heard}}]{maartens2000chaotic}
\bibinfo{author}{\bibfnamefont{R.}~\bibnamefont{Maartens}},
  \bibinfo{author}{\bibfnamefont{D.}~\bibnamefont{Wands}},
  \bibinfo{author}{\bibfnamefont{B.~A.} \bibnamefont{Bassett}},
  \bibnamefont{and} \bibinfo{author}{\bibfnamefont{I.~P.} \bibnamefont{Heard}},
  \bibinfo{journal}{Physical Review D} \textbf{\bibinfo{volume}{62}},
  \bibinfo{pages}{041301} (\bibinfo{year}{2000}).

\bibitem[{\citenamefont{Golanbari et~al.}(2014)\citenamefont{Golanbari,
  Mohammadi, and Saaidi}}]{golanbari2014brane}
\bibinfo{author}{\bibfnamefont{T.}~\bibnamefont{Golanbari}},
  \bibinfo{author}{\bibfnamefont{A.}~\bibnamefont{Mohammadi}},
  \bibnamefont{and} \bibinfo{author}{\bibfnamefont{K.}~\bibnamefont{Saaidi}},
  \bibinfo{journal}{Physical Review D} \textbf{\bibinfo{volume}{89}},
  \bibinfo{pages}{103529} (\bibinfo{year}{2014}).

\bibitem[{\citenamefont{{Mohammadi, Abolhassan and Golanbari, Tayeb and Nasri,
  Salah and Saaidi, Khaled}}(2020)}]{Mohammadi:2020ftb}
\bibinfo{author}{\bibnamefont{{Mohammadi, Abolhassan and Golanbari, Tayeb and
  Nasri, Salah and Saaidi, Khaled}}}, \bibinfo{journal}{Phys. Rev. D}
  \textbf{\bibinfo{volume}{101}}, \bibinfo{pages}{123537}
  (\bibinfo{year}{2020}), \eprint{2004.12137}.

\bibitem[{\citenamefont{Banerjee and Paul}(2017)}]{Banerjee:2017lxi}
\bibinfo{author}{\bibfnamefont{N.}~\bibnamefont{Banerjee}} \bibnamefont{and}
  \bibinfo{author}{\bibfnamefont{T.}~\bibnamefont{Paul}},
  \bibinfo{journal}{Eur. Phys. J. C} \textbf{\bibinfo{volume}{77}},
  \bibinfo{pages}{672} (\bibinfo{year}{2017}), \eprint{arXiv:1706.05964}.

\bibitem[{\citenamefont{Elizalde et~al.}(2019)\citenamefont{Elizalde, Odintsov,
  Paul, and Sáez-Chillón~Gómez}}]{Elizalde:2018rmz}
\bibinfo{author}{\bibfnamefont{E.}~\bibnamefont{Elizalde}},
  \bibinfo{author}{\bibfnamefont{S.~D.} \bibnamefont{Odintsov}},
  \bibinfo{author}{\bibfnamefont{T.}~\bibnamefont{Paul}}, \bibnamefont{and}
  \bibinfo{author}{\bibfnamefont{D.}~\bibnamefont{Sáez-Chillón~Gómez}},
  \bibinfo{journal}{Phys. Rev. D} \textbf{\bibinfo{volume}{99}},
  \bibinfo{pages}{063506} (\bibinfo{year}{2019}), \eprint{arXiv:1811.02960}.

\bibitem[{\citenamefont{Paul and SenGupta}(2019)}]{Paul:2018jpq}
\bibinfo{author}{\bibfnamefont{T.}~\bibnamefont{Paul}} \bibnamefont{and}
  \bibinfo{author}{\bibfnamefont{S.}~\bibnamefont{SenGupta}},
  \bibinfo{journal}{Eur. Phys. J. C} \textbf{\bibinfo{volume}{79}},
  \bibinfo{pages}{591} (\bibinfo{year}{2019}), \eprint{arXiv:1808.00172}.

\bibitem[{\citenamefont{Obied et~al.}(2018)\citenamefont{Obied, Ooguri,
  Spodyneiko, and Vafa}}]{Obied:2018sgi}
\bibinfo{author}{\bibfnamefont{G.}~\bibnamefont{Obied}},
  \bibinfo{author}{\bibfnamefont{H.}~\bibnamefont{Ooguri}},
  \bibinfo{author}{\bibfnamefont{L.}~\bibnamefont{Spodyneiko}},
  \bibnamefont{and} \bibinfo{author}{\bibfnamefont{C.}~\bibnamefont{Vafa}}
  (\bibinfo{year}{2018}), \eprint{arXiv:1806.08362}.

\bibitem[{\citenamefont{Garg and Krishnan}(2019)}]{Garg:2018reu}
\bibinfo{author}{\bibfnamefont{S.~K.} \bibnamefont{Garg}} \bibnamefont{and}
  \bibinfo{author}{\bibfnamefont{C.}~\bibnamefont{Krishnan}},
  \bibinfo{journal}{JHEP} \textbf{\bibinfo{volume}{11}}, \bibinfo{pages}{075}
  (\bibinfo{year}{2019}), \eprint{arXiv:1807.05193}.

\bibitem[{\citenamefont{Ooguri et~al.}(2019)\citenamefont{Ooguri, Palti, Shiu,
  and Vafa}}]{Ooguri:2018wrx}
\bibinfo{author}{\bibfnamefont{H.}~\bibnamefont{Ooguri}},
  \bibinfo{author}{\bibfnamefont{E.}~\bibnamefont{Palti}},
  \bibinfo{author}{\bibfnamefont{G.}~\bibnamefont{Shiu}}, \bibnamefont{and}
  \bibinfo{author}{\bibfnamefont{C.}~\bibnamefont{Vafa}},
  \bibinfo{journal}{Phys. Lett.} \textbf{\bibinfo{volume}{B788}},
  \bibinfo{pages}{180} (\bibinfo{year}{2019}), \eprint{arXiv:1810.05506}.

\bibitem[{\citenamefont{Kehagias and Riotto}(2018)}]{Kehagias:2018uem}
\bibinfo{author}{\bibfnamefont{A.}~\bibnamefont{Kehagias}} \bibnamefont{and}
  \bibinfo{author}{\bibfnamefont{A.}~\bibnamefont{Riotto}},
  \bibinfo{journal}{Fortsch. Phys.} \textbf{\bibinfo{volume}{66}},
  \bibinfo{pages}{1800052} (\bibinfo{year}{2018}), \eprint{arXiv:1807.05445}.

\bibitem[{\citenamefont{Das}(2019)}]{Das:2018rpg}
\bibinfo{author}{\bibfnamefont{S.}~\bibnamefont{Das}}, \bibinfo{journal}{Phys.
  Rev.} \textbf{\bibinfo{volume}{D99}}, \bibinfo{pages}{063514}
  (\bibinfo{year}{2019}), \eprint{arXiv:1810.05038}.

\bibitem[{\citenamefont{Kinney}(2019)}]{Kinney:2018kew}
\bibinfo{author}{\bibfnamefont{W.~H.} \bibnamefont{Kinney}},
  \bibinfo{journal}{Phys. Rev. Lett.} \textbf{\bibinfo{volume}{122}},
  \bibinfo{pages}{081302} (\bibinfo{year}{2019}), \eprint{arXiv:1811.11698}.

\bibitem[{\citenamefont{Matsui and Takahashi}(2019)}]{Matsui:2018bsy}
\bibinfo{author}{\bibfnamefont{H.}~\bibnamefont{Matsui}} \bibnamefont{and}
  \bibinfo{author}{\bibfnamefont{F.}~\bibnamefont{Takahashi}},
  \bibinfo{journal}{Phys. Rev.} \textbf{\bibinfo{volume}{D99}},
  \bibinfo{pages}{023533} (\bibinfo{year}{2019}), \eprint{arXiv:1807.11938}.

\bibitem[{\citenamefont{Lin}(2019)}]{Lin:2018rnx}
\bibinfo{author}{\bibfnamefont{C.-M.} \bibnamefont{Lin}},
  \bibinfo{journal}{Phys. Rev.} \textbf{\bibinfo{volume}{D99}},
  \bibinfo{pages}{023519} (\bibinfo{year}{2019}), \eprint{arXiv:1810.11992}.

\bibitem[{\citenamefont{Dimopoulos}(2018)}]{Dimopoulos:2018upl}
\bibinfo{author}{\bibfnamefont{K.}~\bibnamefont{Dimopoulos}},
  \bibinfo{journal}{Phys. Rev.} \textbf{\bibinfo{volume}{D98}},
  \bibinfo{pages}{123516} (\bibinfo{year}{2018}), \eprint{arXiv:1810.03438}.

\bibitem[{\citenamefont{Kinney et~al.}(2019)\citenamefont{Kinney, Vagnozzi, and
  Visinelli}}]{Kinney:2018nny}
\bibinfo{author}{\bibfnamefont{W.~H.} \bibnamefont{Kinney}},
  \bibinfo{author}{\bibfnamefont{S.}~\bibnamefont{Vagnozzi}}, \bibnamefont{and}
  \bibinfo{author}{\bibfnamefont{L.}~\bibnamefont{Visinelli}},
  \bibinfo{journal}{Class. Quant. Grav.} \textbf{\bibinfo{volume}{36}},
  \bibinfo{pages}{117001} (\bibinfo{year}{2019}), \eprint{arXiv:1808.06424}.

\bibitem[{\citenamefont{Geng}(2019)}]{Geng:2019phi}
\bibinfo{author}{\bibfnamefont{H.}~\bibnamefont{Geng}} (\bibinfo{year}{2019}),
  \eprint{arXiv:1910.14047}.

\bibitem[{\citenamefont{Brahma and Wali~Hossain}(2019)}]{Brahma:2018hrd}
\bibinfo{author}{\bibfnamefont{S.}~\bibnamefont{Brahma}} \bibnamefont{and}
  \bibinfo{author}{\bibfnamefont{M.}~\bibnamefont{Wali~Hossain}},
  \bibinfo{journal}{JHEP} \textbf{\bibinfo{volume}{03}}, \bibinfo{pages}{006}
  (\bibinfo{year}{2019}), \eprint{arXiv:1809.01277}.

\bibitem[{\citenamefont{Brahma and Shandera}(2019)}]{Brahma:2019iyy}
\bibinfo{author}{\bibfnamefont{S.}~\bibnamefont{Brahma}} \bibnamefont{and}
  \bibinfo{author}{\bibfnamefont{S.}~\bibnamefont{Shandera}},
  \bibinfo{journal}{JHEP} \textbf{\bibinfo{volume}{11}}, \bibinfo{pages}{016}
  (\bibinfo{year}{2019}), \eprint{arXiv:1904.10979}.

\bibitem[{\citenamefont{Wang et~al.}(2019)\citenamefont{Wang, Brandenberger,
  and Heisenberg}}]{Wang:2019eym}
\bibinfo{author}{\bibfnamefont{Z.}~\bibnamefont{Wang}},
  \bibinfo{author}{\bibfnamefont{R.}~\bibnamefont{Brandenberger}},
  \bibnamefont{and}
  \bibinfo{author}{\bibfnamefont{L.}~\bibnamefont{Heisenberg}}
  (\bibinfo{year}{2019}), \eprint{arXiv:1907.08943}.

\bibitem[{\citenamefont{Odintsov and Oikonomou}(2020)}]{Odintsov:2020zkl}
\bibinfo{author}{\bibfnamefont{S.}~\bibnamefont{Odintsov}} \bibnamefont{and}
  \bibinfo{author}{\bibfnamefont{V.}~\bibnamefont{Oikonomou}},
  \bibinfo{journal}{Phys. Lett. B} \textbf{\bibinfo{volume}{805}},
  \bibinfo{pages}{135437} (\bibinfo{year}{2020}), \eprint{arXiv:2004.00479}.

\bibitem[{\citenamefont{Odintsov and Oikonomou}(2019)}]{Odintsov:2018zai}
\bibinfo{author}{\bibfnamefont{S.}~\bibnamefont{Odintsov}} \bibnamefont{and}
  \bibinfo{author}{\bibfnamefont{V.}~\bibnamefont{Oikonomou}},
  \bibinfo{journal}{EPL} \textbf{\bibinfo{volume}{126}}, \bibinfo{pages}{20002}
  (\bibinfo{year}{2019}), \eprint{arXiv:1810.03575}.

\bibitem[{\citenamefont{Mohammadi
  et~al.}(2020{\natexlab{a}})\citenamefont{Mohammadi, Golanbari, Sheikhahmadi,
  Sayar, Akhtari, Rasheed, and Saaidi}}]{Rasheed:2020syk}
\bibinfo{author}{\bibfnamefont{A.}~\bibnamefont{Mohammadi}},
  \bibinfo{author}{\bibfnamefont{T.}~\bibnamefont{Golanbari}},
  \bibinfo{author}{\bibfnamefont{H.}~\bibnamefont{Sheikhahmadi}},
  \bibinfo{author}{\bibfnamefont{K.}~\bibnamefont{Sayar}},
  \bibinfo{author}{\bibfnamefont{L.}~\bibnamefont{Akhtari}},
  \bibinfo{author}{\bibfnamefont{M.}~\bibnamefont{Rasheed}}, \bibnamefont{and}
  \bibinfo{author}{\bibfnamefont{K.}~\bibnamefont{Saaidi}}
  (\bibinfo{year}{2020}{\natexlab{a}}), \eprint{arXiv:2001.10042}.

\bibitem[{\citenamefont{Brahma et~al.}(2020)\citenamefont{Brahma,
  Brandenberger, and Yeom}}]{Brahma:2020cpy}
\bibinfo{author}{\bibfnamefont{S.}~\bibnamefont{Brahma}},
  \bibinfo{author}{\bibfnamefont{R.}~\bibnamefont{Brandenberger}},
  \bibnamefont{and} \bibinfo{author}{\bibfnamefont{D.-H.} \bibnamefont{Yeom}}
  (\bibinfo{year}{2020}), \eprint{arXiv:2002.02941}.

\bibitem[{\citenamefont{Bedroya and Vafa}(2020)}]{Bedroya:2019snp}
\bibinfo{author}{\bibfnamefont{A.}~\bibnamefont{Bedroya}} \bibnamefont{and}
  \bibinfo{author}{\bibfnamefont{C.}~\bibnamefont{Vafa}},
  \bibinfo{journal}{JHEP} \textbf{\bibinfo{volume}{09}}, \bibinfo{pages}{123}
  (\bibinfo{year}{2020}), \eprint{1909.11063}.

\bibitem[{\citenamefont{Bedroya et~al.}(2020)\citenamefont{Bedroya,
  Brandenberger, Loverde, and Vafa}}]{Bedroya:2019tba}
\bibinfo{author}{\bibfnamefont{A.}~\bibnamefont{Bedroya}},
  \bibinfo{author}{\bibfnamefont{R.}~\bibnamefont{Brandenberger}},
  \bibinfo{author}{\bibfnamefont{M.}~\bibnamefont{Loverde}}, \bibnamefont{and}
  \bibinfo{author}{\bibfnamefont{C.}~\bibnamefont{Vafa}},
  \bibinfo{journal}{Phys. Rev. D} \textbf{\bibinfo{volume}{101}},
  \bibinfo{pages}{103502} (\bibinfo{year}{2020}), \eprint{1909.11106}.

\bibitem[{\citenamefont{Mohammadi
  et~al.}(2020{\natexlab{b}})\citenamefont{Mohammadi, Golanbari, and
  Enayati}}]{Mohammadi:2020ctd}
\bibinfo{author}{\bibfnamefont{A.}~\bibnamefont{Mohammadi}},
  \bibinfo{author}{\bibfnamefont{T.}~\bibnamefont{Golanbari}},
  \bibnamefont{and} \bibinfo{author}{\bibfnamefont{J.}~\bibnamefont{Enayati}}
  (\bibinfo{year}{2020}{\natexlab{b}}), \eprint{2012.01512}.

\bibitem[{\citenamefont{Cline et~al.}(1999)\citenamefont{Cline, Grojean, and
  Servant}}]{Cline:1999ts}
\bibinfo{author}{\bibfnamefont{J.~M.} \bibnamefont{Cline}},
  \bibinfo{author}{\bibfnamefont{C.}~\bibnamefont{Grojean}}, \bibnamefont{and}
  \bibinfo{author}{\bibfnamefont{G.}~\bibnamefont{Servant}},
  \bibinfo{journal}{Phys. Rev. Lett.} \textbf{\bibinfo{volume}{83}},
  \bibinfo{pages}{4245} (\bibinfo{year}{1999}), \eprint{hep-ph/9906523}.

\bibitem[{\citenamefont{Germani and Maartens}(2001)}]{Germani:2001du}
\bibinfo{author}{\bibfnamefont{C.}~\bibnamefont{Germani}} \bibnamefont{and}
  \bibinfo{author}{\bibfnamefont{R.}~\bibnamefont{Maartens}},
  \bibinfo{journal}{Phys. Rev.} \textbf{\bibinfo{volume}{D64}},
  \bibinfo{pages}{124010} (\bibinfo{year}{2001}), \eprint{hep-th/0107011}.

\bibitem[{\citenamefont{Wands et~al.}(2000)\citenamefont{Wands, Malik, Lyth,
  and Liddle}}]{Wands:2000dp}
\bibinfo{author}{\bibfnamefont{D.}~\bibnamefont{Wands}},
  \bibinfo{author}{\bibfnamefont{K.~A.} \bibnamefont{Malik}},
  \bibinfo{author}{\bibfnamefont{D.~H.} \bibnamefont{Lyth}}, \bibnamefont{and}
  \bibinfo{author}{\bibfnamefont{A.~R.} \bibnamefont{Liddle}},
  \bibinfo{journal}{Phys. Rev.} \textbf{\bibinfo{volume}{D62}},
  \bibinfo{pages}{043527} (\bibinfo{year}{2000}), \eprint{astro-ph/0003278}.

\bibitem[{\citenamefont{Brax et~al.}(2004)\citenamefont{Brax, van~de Bruck, and
  Davis}}]{Brax:2004xh}
\bibinfo{author}{\bibfnamefont{P.}~\bibnamefont{Brax}},
  \bibinfo{author}{\bibfnamefont{C.}~\bibnamefont{van~de Bruck}},
  \bibnamefont{and} \bibinfo{author}{\bibfnamefont{A.-C.} \bibnamefont{Davis}},
  \bibinfo{journal}{Rept. Prog. Phys.} \textbf{\bibinfo{volume}{67}},
  \bibinfo{pages}{2183} (\bibinfo{year}{2004}), \eprint{hep-th/0404011}.

\bibitem[{\citenamefont{Langlois et~al.}(2000)\citenamefont{Langlois, Maartens,
  and Wands}}]{Langlois:2000ns}
\bibinfo{author}{\bibfnamefont{D.}~\bibnamefont{Langlois}},
  \bibinfo{author}{\bibfnamefont{R.}~\bibnamefont{Maartens}}, \bibnamefont{and}
  \bibinfo{author}{\bibfnamefont{D.}~\bibnamefont{Wands}},
  \bibinfo{journal}{Phys. Lett.} \textbf{\bibinfo{volume}{B489}},
  \bibinfo{pages}{259} (\bibinfo{year}{2000}), \eprint{hep-th/0006007}.

\bibitem[{\citenamefont{Huey and Lidsey}(2001)}]{Huey:2001ae}
\bibinfo{author}{\bibfnamefont{G.}~\bibnamefont{Huey}} \bibnamefont{and}
  \bibinfo{author}{\bibfnamefont{J.~E.} \bibnamefont{Lidsey}},
  \bibinfo{journal}{Phys. Lett.} \textbf{\bibinfo{volume}{B514}},
  \bibinfo{pages}{217} (\bibinfo{year}{2001}), \eprint{astro-ph/0104006}.

\bibitem[{\citenamefont{Lin et~al.}(2019)\citenamefont{Lin, Ng, and
  Cheung}}]{Lin:2018kjm}
\bibinfo{author}{\bibfnamefont{C.-M.} \bibnamefont{Lin}},
  \bibinfo{author}{\bibfnamefont{K.-W.} \bibnamefont{Ng}}, \bibnamefont{and}
  \bibinfo{author}{\bibfnamefont{K.}~\bibnamefont{Cheung}},
  \bibinfo{journal}{Phys. Rev. D} \textbf{\bibinfo{volume}{100}},
  \bibinfo{pages}{023545} (\bibinfo{year}{2019}), \eprint{1810.01644}.

\bibitem[{\citenamefont{Adams et~al.}(1993)\citenamefont{Adams, Bond, Freese,
  Frieman, and Olinto}}]{Adams:1992bn}
\bibinfo{author}{\bibfnamefont{F.~C.} \bibnamefont{Adams}},
  \bibinfo{author}{\bibfnamefont{J.}~\bibnamefont{Bond}},
  \bibinfo{author}{\bibfnamefont{K.}~\bibnamefont{Freese}},
  \bibinfo{author}{\bibfnamefont{J.~A.} \bibnamefont{Frieman}},
  \bibnamefont{and} \bibinfo{author}{\bibfnamefont{A.~V.}
  \bibnamefont{Olinto}}, \bibinfo{journal}{Phys. Rev. D}
  \textbf{\bibinfo{volume}{47}}, \bibinfo{pages}{426} (\bibinfo{year}{1993}),
  \eprint{hep-ph/9207245}.

\bibitem[{\citenamefont{Kallosh et~al.}(2013)\citenamefont{Kallosh, Linde, and
  Roest}}]{Kallosh:2013yoa}
\bibinfo{author}{\bibfnamefont{R.}~\bibnamefont{Kallosh}},
  \bibinfo{author}{\bibfnamefont{A.}~\bibnamefont{Linde}}, \bibnamefont{and}
  \bibinfo{author}{\bibfnamefont{D.}~\bibnamefont{Roest}},
  \bibinfo{journal}{JHEP} \textbf{\bibinfo{volume}{11}}, \bibinfo{pages}{198}
  (\bibinfo{year}{2013}), \eprint{1311.0472}.

\bibitem[{\citenamefont{Kallosh et~al.}(2014)\citenamefont{Kallosh, Linde, and
  Roest}}]{Kallosh:2013tua}
\bibinfo{author}{\bibfnamefont{R.}~\bibnamefont{Kallosh}},
  \bibinfo{author}{\bibfnamefont{A.}~\bibnamefont{Linde}}, \bibnamefont{and}
  \bibinfo{author}{\bibfnamefont{D.}~\bibnamefont{Roest}},
  \bibinfo{journal}{Phys. Rev. Lett.} \textbf{\bibinfo{volume}{112}},
  \bibinfo{pages}{011303} (\bibinfo{year}{2014}), \eprint{1310.3950}.

\bibitem[{\citenamefont{Ferrara
  et~al.}(2013{\natexlab{a}})\citenamefont{Ferrara, Kallosh, Linde, and
  Porrati}}]{Ferrara:2013rsa}
\bibinfo{author}{\bibfnamefont{S.}~\bibnamefont{Ferrara}},
  \bibinfo{author}{\bibfnamefont{R.}~\bibnamefont{Kallosh}},
  \bibinfo{author}{\bibfnamefont{A.}~\bibnamefont{Linde}}, \bibnamefont{and}
  \bibinfo{author}{\bibfnamefont{M.}~\bibnamefont{Porrati}},
  \bibinfo{journal}{Phys. Rev. D} \textbf{\bibinfo{volume}{88}},
  \bibinfo{pages}{085038} (\bibinfo{year}{2013}{\natexlab{a}}),
  \eprint{1307.7696}.

\bibitem[{\citenamefont{Ferrara
  et~al.}(2013{\natexlab{b}})\citenamefont{Ferrara, Kallosh, Linde, and
  Porrati}}]{Ferrara:2013kca}
\bibinfo{author}{\bibfnamefont{S.}~\bibnamefont{Ferrara}},
  \bibinfo{author}{\bibfnamefont{R.}~\bibnamefont{Kallosh}},
  \bibinfo{author}{\bibfnamefont{A.}~\bibnamefont{Linde}}, \bibnamefont{and}
  \bibinfo{author}{\bibfnamefont{M.}~\bibnamefont{Porrati}},
  \bibinfo{journal}{JCAP} \textbf{\bibinfo{volume}{11}}, \bibinfo{pages}{046}
  (\bibinfo{year}{2013}{\natexlab{b}}), \eprint{1309.1085}.

\bibitem[{\citenamefont{Dimopoulos and Owen}(2017)}]{Dimopoulos:2017zvq}
\bibinfo{author}{\bibfnamefont{K.}~\bibnamefont{Dimopoulos}} \bibnamefont{and}
  \bibinfo{author}{\bibfnamefont{C.}~\bibnamefont{Owen}},
  \bibinfo{journal}{JCAP} \textbf{\bibinfo{volume}{06}}, \bibinfo{pages}{027}
  (\bibinfo{year}{2017}), \eprint{1703.00305}.

\bibitem[{\citenamefont{Ueno and Yamamoto}(2016)}]{Ueno:2016dim}
\bibinfo{author}{\bibfnamefont{Y.}~\bibnamefont{Ueno}} \bibnamefont{and}
  \bibinfo{author}{\bibfnamefont{K.}~\bibnamefont{Yamamoto}},
  \bibinfo{journal}{Phys. Rev. D} \textbf{\bibinfo{volume}{93}},
  \bibinfo{pages}{083524} (\bibinfo{year}{2016}), \eprint{1602.07427}.

\bibitem[{\citenamefont{Liddle and Leach}(2003)}]{Liddle:2003as}
\bibinfo{author}{\bibfnamefont{A.~R.} \bibnamefont{Liddle}} \bibnamefont{and}
  \bibinfo{author}{\bibfnamefont{S.~M.} \bibnamefont{Leach}},
  \bibinfo{journal}{Phys. Rev. D} \textbf{\bibinfo{volume}{68}},
  \bibinfo{pages}{103503} (\bibinfo{year}{2003}), \eprint{astro-ph/0305263}.

\bibitem[{\citenamefont{Dai et~al.}(2014)\citenamefont{Dai, Kamionkowski, and
  Wang}}]{Dai:2014jja}
\bibinfo{author}{\bibfnamefont{L.}~\bibnamefont{Dai}},
  \bibinfo{author}{\bibfnamefont{M.}~\bibnamefont{Kamionkowski}},
  \bibnamefont{and} \bibinfo{author}{\bibfnamefont{J.}~\bibnamefont{Wang}},
  \bibinfo{journal}{Phys. Rev. Lett.} \textbf{\bibinfo{volume}{113}},
  \bibinfo{pages}{041302} (\bibinfo{year}{2014}), \eprint{1404.6704}.

\bibitem[{\citenamefont{Cook et~al.}(2015)\citenamefont{Cook, Dimastrogiovanni,
  Easson, and Krauss}}]{Cook:2015vqa}
\bibinfo{author}{\bibfnamefont{J.~L.} \bibnamefont{Cook}},
  \bibinfo{author}{\bibfnamefont{E.}~\bibnamefont{Dimastrogiovanni}},
  \bibinfo{author}{\bibfnamefont{D.~A.} \bibnamefont{Easson}},
  \bibnamefont{and} \bibinfo{author}{\bibfnamefont{L.~M.}
  \bibnamefont{Krauss}}, \bibinfo{journal}{JCAP} \textbf{\bibinfo{volume}{04}},
  \bibinfo{pages}{047} (\bibinfo{year}{2015}), \eprint{1502.04673}.

\bibitem[{\citenamefont{Bhattacharjee et~al.}(2017)\citenamefont{Bhattacharjee,
  Maity, and Mukherjee}}]{Bhattacharjee:2016ohe}
\bibinfo{author}{\bibfnamefont{S.}~\bibnamefont{Bhattacharjee}},
  \bibinfo{author}{\bibfnamefont{D.}~\bibnamefont{Maity}}, \bibnamefont{and}
  \bibinfo{author}{\bibfnamefont{R.}~\bibnamefont{Mukherjee}},
  \bibinfo{journal}{Phys. Rev. D} \textbf{\bibinfo{volume}{95}},
  \bibinfo{pages}{023514} (\bibinfo{year}{2017}), \eprint{1606.00698}.

\bibitem[{\citenamefont{Drewes et~al.}(2017)\citenamefont{Drewes, Kang, and
  Mun}}]{Drewes:2017fmn}
\bibinfo{author}{\bibfnamefont{M.}~\bibnamefont{Drewes}},
  \bibinfo{author}{\bibfnamefont{J.~U.} \bibnamefont{Kang}}, \bibnamefont{and}
  \bibinfo{author}{\bibfnamefont{U.~R.} \bibnamefont{Mun}},
  \bibinfo{journal}{JHEP} \textbf{\bibinfo{volume}{11}}, \bibinfo{pages}{072}
  (\bibinfo{year}{2017}), \eprint{1708.01197}.

\bibitem[{\citenamefont{McAllister et~al.}(2010)\citenamefont{McAllister,
  Silverstein, and Westphal}}]{McAllister:2008hb}
\bibinfo{author}{\bibfnamefont{L.}~\bibnamefont{McAllister}},
  \bibinfo{author}{\bibfnamefont{E.}~\bibnamefont{Silverstein}},
  \bibnamefont{and} \bibinfo{author}{\bibfnamefont{A.}~\bibnamefont{Westphal}},
  \bibinfo{journal}{Phys. Rev. D} \textbf{\bibinfo{volume}{82}},
  \bibinfo{pages}{046003} (\bibinfo{year}{2010}), \eprint{0808.0706}.

\bibitem[{\citenamefont{Silverstein and Westphal}(2008)}]{Silverstein:2008sg}
\bibinfo{author}{\bibfnamefont{E.}~\bibnamefont{Silverstein}} \bibnamefont{and}
  \bibinfo{author}{\bibfnamefont{A.}~\bibnamefont{Westphal}},
  \bibinfo{journal}{Phys. Rev. D} \textbf{\bibinfo{volume}{78}},
  \bibinfo{pages}{106003} (\bibinfo{year}{2008}), \eprint{0803.3085}.

\bibitem[{\citenamefont{Podolsky et~al.}(2006)\citenamefont{Podolsky, Felder,
  Kofman, and Peloso}}]{Podolsky:2005bw}
\bibinfo{author}{\bibfnamefont{D.~I.} \bibnamefont{Podolsky}},
  \bibinfo{author}{\bibfnamefont{G.~N.} \bibnamefont{Felder}},
  \bibinfo{author}{\bibfnamefont{L.}~\bibnamefont{Kofman}}, \bibnamefont{and}
  \bibinfo{author}{\bibfnamefont{M.}~\bibnamefont{Peloso}},
  \bibinfo{journal}{Phys. Rev. D} \textbf{\bibinfo{volume}{73}},
  \bibinfo{pages}{023501} (\bibinfo{year}{2006}), \eprint{hep-ph/0507096}.

\bibitem[{\citenamefont{Martin and Ringeval}(2010)}]{Martin:2010kz}
\bibinfo{author}{\bibfnamefont{J.}~\bibnamefont{Martin}} \bibnamefont{and}
  \bibinfo{author}{\bibfnamefont{C.}~\bibnamefont{Ringeval}},
  \bibinfo{journal}{Phys. Rev. D} \textbf{\bibinfo{volume}{82}},
  \bibinfo{pages}{023511} (\bibinfo{year}{2010}), \eprint{1004.5525}.

\bibitem[{\citenamefont{Mohammadi
  et~al.}(2020{\natexlab{c}})\citenamefont{Mohammadi, Golanbari, Enayati,
  Jalalzadeh, and Saaidi}}]{Mohammadi:2020twg}
\bibinfo{author}{\bibfnamefont{A.}~\bibnamefont{Mohammadi}},
  \bibinfo{author}{\bibfnamefont{T.}~\bibnamefont{Golanbari}},
  \bibinfo{author}{\bibfnamefont{J.}~\bibnamefont{Enayati}},
  \bibinfo{author}{\bibfnamefont{S.}~\bibnamefont{Jalalzadeh}},
  \bibnamefont{and} \bibinfo{author}{\bibfnamefont{K.}~\bibnamefont{Saaidi}}
  (\bibinfo{year}{2020}{\natexlab{c}}), \eprint{2011.13957}.

\bibitem[{\citenamefont{Lozanov}(2019)}]{Lozanov:2019jxc}
\bibinfo{author}{\bibfnamefont{K.~D.} \bibnamefont{Lozanov}}
  (\bibinfo{year}{2019}), \eprint{1907.04402}.

\bibitem[{\citenamefont{Allahverdi}(2000)}]{Allahverdi:2000ss}
\bibinfo{author}{\bibfnamefont{R.}~\bibnamefont{Allahverdi}},
  \bibinfo{journal}{Phys. Rev. D} \textbf{\bibinfo{volume}{62}},
  \bibinfo{pages}{063509} (\bibinfo{year}{2000}), \eprint{hep-ph/0004035}.

\end{thebibliography}

%\end{thebibliography}

%%%%%%%%%%%%%%%%%%%%%%%%%%%%%%%%%%%%%%%%%%%%%%%%%%%%%%%%%%%%%%%%%%%%%
%%%%%%%%%%%%%%%%%%%%%%%%%%%%%%%%%%%%%%%%%%%%%%%%%%%%%%%%%%%%%%%%%%%%

\end{document}